%% file: CGL.tex
\documentclass[english,nofootinbib,aps,pra,reprint,10pt]{revtex4-1}
\usepackage[T1]{fontenc}
\usepackage[latin9]{inputenc}
\usepackage{amsmath}
\usepackage{amssymb}
\usepackage{graphicx}

\makeatletter

\providecommand{\tabularnewline}{\\}

\makeatother

\usepackage{babel}
\begin{document}
\input{MathMacros.tex}

\title{Transition to classical regime in quantum mechanics on a lattice \\
and implications of discontinuous space}
\author{Oleg Kabernik}
\email{ok1223@gmail.com}

\affiliation{Department of Physics and Astronomy, University of British Columbia,
Vancouver, BC, V6T 1Z1, Canada}
\date{\today}
\begin{abstract}
It is well known that due to the uncertainty principle the Planck
constant sets a resolution boundary in phase space and the resulting
trade-off in resolutions between incompatible measurements has been
thoroughly investigated. It is also known that in the classical regime
sufficiently coarse measurements of position and momentum can simultaneously
be determined. However, the picture of how the uncertainty principle
gradually disappears as we transition from the quantum to the classical
regime is not so vivid. In the present work we will clarify this picture
by studying the associated probabilities that quantify the effects
of the uncertainty principle in the framework of finite-dimensional
quantum mechanics on a lattice. We will also study how these probabilities
are perturbed by the granularity of the lattice and show that they
can signal the discontinuity of the underlying space.
\end{abstract}
\maketitle

\section{Introduction}

Heisenberg\textquoteright s uncertainty principle is colloquially
understood as the fact that arbitrarily precise values of position
and momentum cannot simultaneously be determined (see \citep{busch2006complementarity,busch2007heisenberg}
for a review). A rigorous formulation of the uncertainty principle
is often conflated with the uncertainty relations for states $\sigma_{x}\sigma_{p}\geq\hbar/2$,
where $\sigma_{x}$ and $\sigma_{p}$ refer to the standard deviations
of independently measured position and momentum of a particle in the
same state. This inequality is also known as \emph{preparation uncertainty
relations} because it rules out the possibility of preparing quantum
states with arbitrarily sharp values of both position and moment.
It does not, however, rule out the possibility of measurements that
simultaneously determine both of these values with arbitrary precision.
The essential effect that rules out the latter possibility is the
mutual disturbance between measurements of incompatible observables,
also known as \emph{error-disturbance uncertainty relations}.

\begin{figure}[t]
\includegraphics[width=0.49\columnwidth]{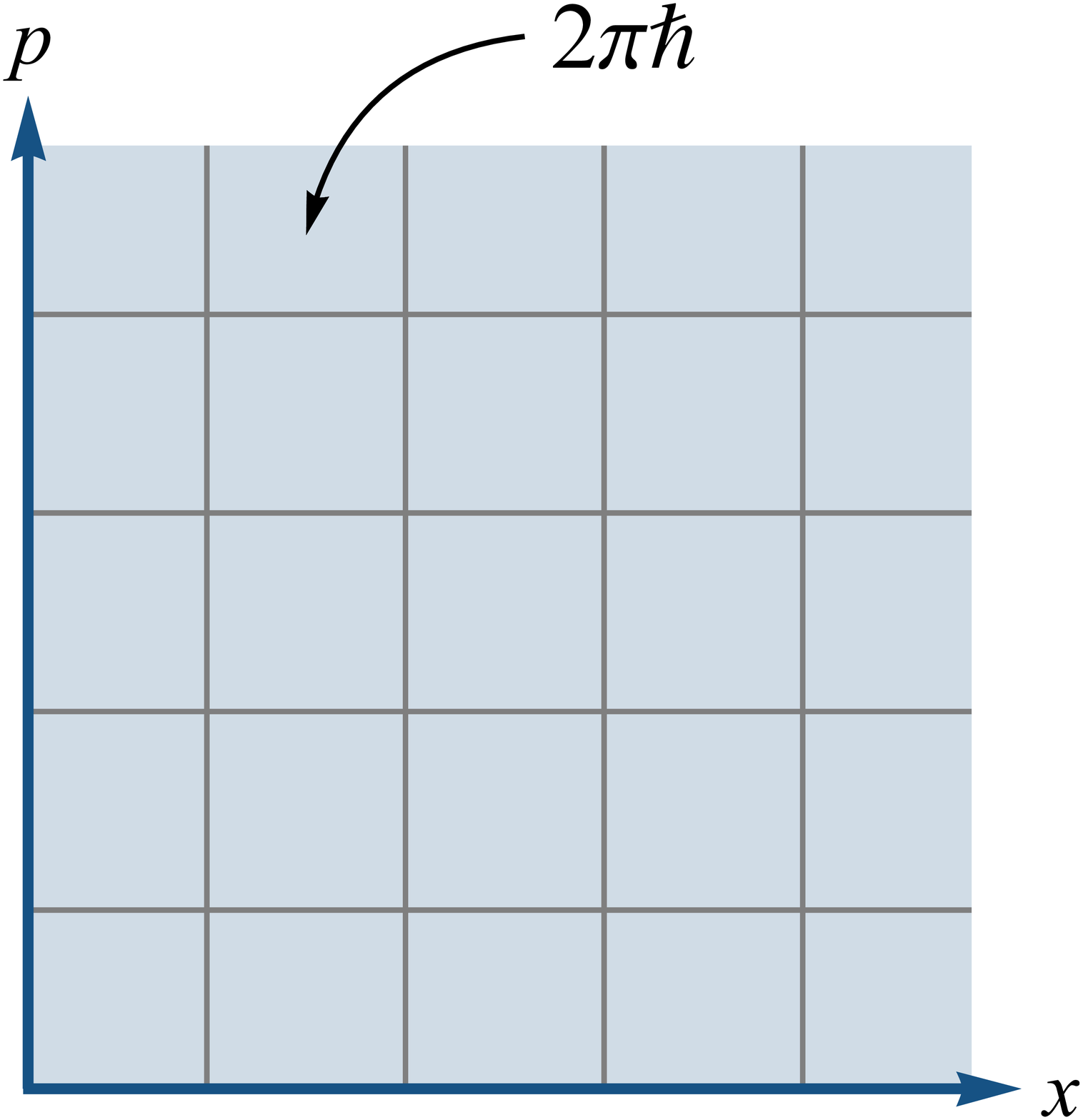} \includegraphics[width=0.49\columnwidth]{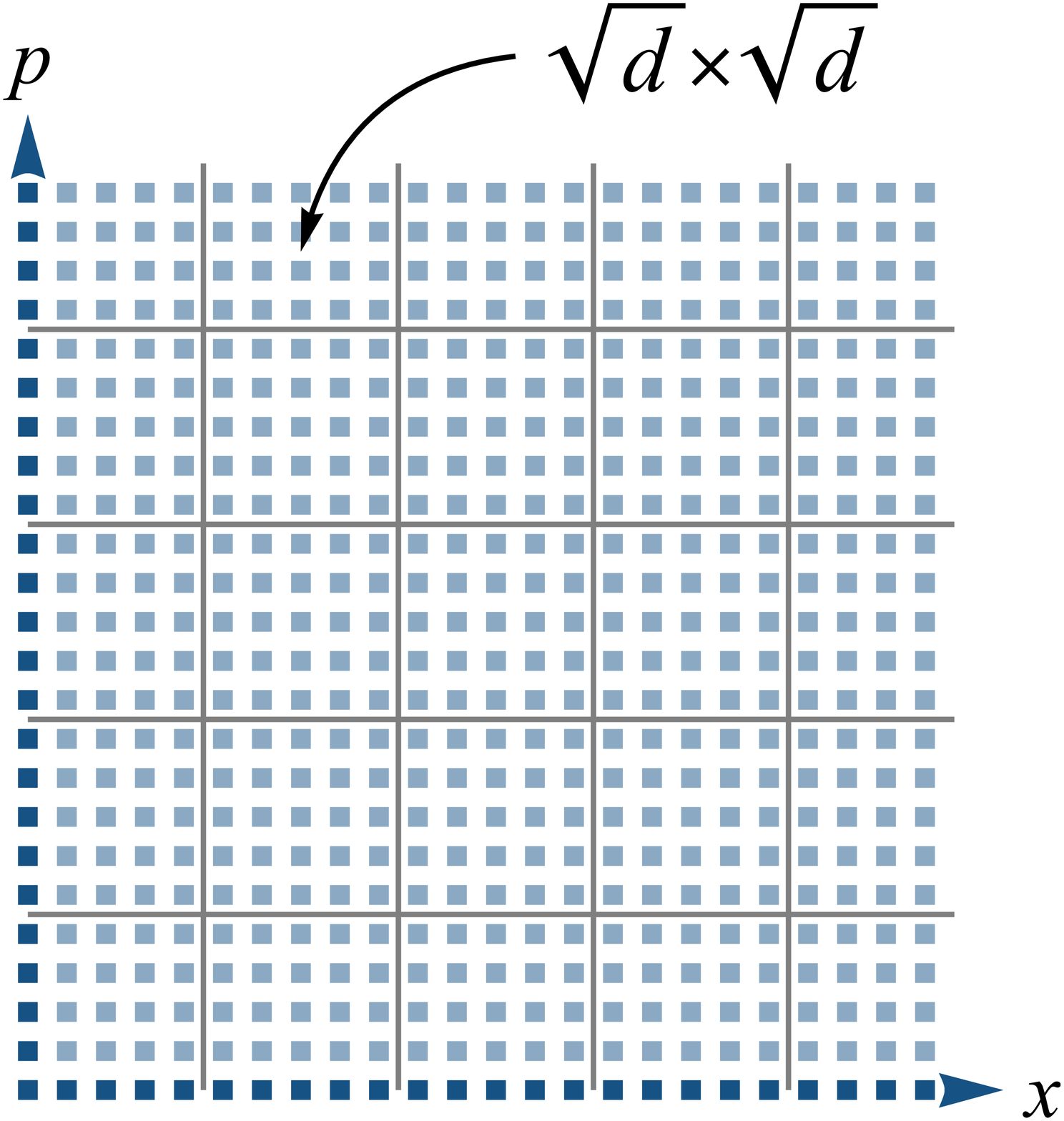}

\caption{\label{fig:phase space diag}(left) The continuous phase space where
the cells with the area $2\pi\hbar$ represent the resolution scale
associated with the uncertainty principle. (right) The discretized
phase space of a lattice of integer length $d$. Analogous cells with
the area $\sqrt{d}\times\sqrt{d}$ arise from the scale $\sqrt{d}$
which is the geometric mean of the minimal length ($1$) and the maximal
length ($d$) on a unitless lattice. The Planck constant $2\pi\hbar$
can be recovered from $\sqrt{d}$ by converting the phase space area
$\sqrt{d}\times\sqrt{d}$ to proper units.}
\end{figure}

According to the original formulation by Heisenberg \citep{heisenberg1927},
due to the unavoidable disturbance by measurements, it is not possible
to localize a particle in a phase space cell of the size of the Planck
constant or smaller. However, when phase space cells much coarser
than the Planck constant are considered, Heisenberg argued that the
values of both observables can be estimated at the expense of lower
resolution. The picture that emerges from Heisenberg's original arguments
is that the Planck constant sets a resolution boundary in phase space
(see Fig. \ref{fig:phase space diag} left) that separates the quantum
scale from the classical scale. There is, of course, a continuum
of scales and it is natural to ask for a characteristic function that
outlines how the uncertainty principle becomes inconsequential as
we decrease the resolution of measurements.

A rigorous formulation of the error-disturbance uncertainty relations
has been extensively debated in recent years \citep{Ozawa03Universally,Busch13proof,branciard2013error,korzekwa2014operational,buscemi2014noise,rozema2015note},
producing multiple perspectives on the fundamental limits of simultaneous
measurability of incompatible observables. These formulations are
similar to the preparation uncertainty relations as they capture the
trade-off between the resolution and disturbance of measurements (which
may also depend on the states). However, the error-disturbance relations
focus on the limits of simultaneous measurability but they do not
outline how the mutual disturbance effects fade away with decreasing
resolution of measurements.

In the present work we will study the mutual disturbance effects of
the uncertainty principle on a finite-dimensional lattice of integer
length $d$. We will quantify the mutual disturbance effects with
the average probability $\left\langle \boldsymbol{p}_{\textrm{agree}}\right\rangle $
that an instantaneous succession of coarse-grained measurements of
position-momentum-position will agree on both outcomes of position.
Since the value $\left\langle \boldsymbol{p}_{\textrm{agree}}\right\rangle $
measures the strength of the mutual disturbance effects as a function
of measurement resolution, it will allow us to quantitatively outline
the transition from the quantum to the classical regime where the
mutual disturbance effects fade away. With that we will show that
the geometric mean of the minimal length and the maximal length on
a lattice is a significant scale that separates the classical regime
of joint measurability, from the quantum regime where mutual disturbance
effects are important (see Fig. \ref{fig:phase space diag} right).

The idea of using coarse-grained measurements to study the quantum-to-classical
transitions is not new. Most notably (and what initially inspired
this work) is the work of Asher Peres \citep{peres2006quantum}, and
later of Kofler and Brukner \citep{kofler2007classical}, where it
was argued that classical physics arises from sufficiently coarse
measurements. This idea has also been investigated from the perspectives
of entanglement observability \citep{Raeisi11} and Bell's or Leggett-Garg
inequalities \citep{Jeong14}. There are also a series of studies
by Rudnicki et al \citep{Rudnicki_2012a,Rudnicki_2012b,toscano2018uncertainty}
on uncertainty relations for coarse-grained observables. What is different
about the present work is that we do not focus on the limits captured
by a certain bound (as in Bell's inequalities or uncertainty relations)
but on the average case captured by the probability $\left\langle \boldsymbol{p}_{\textrm{agree}}\right\rangle $.

Our analysis of the mutual disturbance effects on a lattice with discretized
lengths are also related to what is known as the \emph{generalized
uncertainty principle} \citep{ali2009discreteness}. The idea of the
generalized uncertainty principle follows from the fact that the continuous
phase space picture is incompatible with the various approaches to
quantum gravity \citep{hossenfelder2013minimal} where the minimal
resolvable length is $\delta x\sim10^{-35}\,\textrm{m}$. The existence
of such minimal length should affect the uncertainty principle and
it is usually captured by modifying the canonical commutation relations
\citep{ali2009discreteness}.

There is great interest in identifying observable effects associated
with the modifications of the uncertainty principle due to minimal
length, and in recent years there have been at least two experimental
proposals \citep{ali2011proposal,pikovski2012probing} based on this
idea. Here we will capture the same effect of minimal length, but
instead of modifying the canonical commutation relations we will show
how $\left\langle \boldsymbol{p}_{\textrm{agree}}\right\rangle $
is perturbed by non-vanishing $\delta x$.

\section{From quantum to classical regimes on a lattice}

Let us consider the simple, operationally meaningful quantity $\boldsymbol{p}_{\textrm{agree}}$,
which is the probability that an instantaneous succession of position-momentum-position
measurements will agree on both outcomes of position, regardless of
the outcomes. When all measurements have arbitrarily fine resolution,
the second measurement in this succession prepares a sharp momentum
state that is nearly uniformly distributed in position space. Then,
the probability that the first and the last measurements of position
will agree is vanishingly small $\boldsymbol{p}_{\textrm{agree}}\approx0$.
As we decrease the resolution of measurements, we expect the probability
$\boldsymbol{p}_{\textrm{agree}}$ to grow from $0$ to $1$ because
coarser momentum measurement will cause less spread in the position
space, and coarser position measurements will be more likely to agree
on the estimate of position.

Now, consider the average $\left\langle \boldsymbol{p}_{\textrm{agree}}\right\rangle $
over all states. In general, the average value $\left\langle \boldsymbol{p}_{\textrm{agree}}\right\rangle $
does not inform us about how strongly the measurements disturb each
other for any particular state $\rho$. However, when the average
$\left\langle \boldsymbol{p}_{\textrm{agree}}\right\rangle $ is close
to $0$ or $1$, the value of $\boldsymbol{p}_{\textrm{agree}}\left(\rho\right)$
has to converge to the average for almost all states $\rho$. That
is because $\boldsymbol{p}_{\textrm{agree}}\in\left[0,1\right]$ so
the variance has to vanish as the average gets close to the edges.
Therefore, the value of $\left\langle \boldsymbol{p}_{\textrm{agree}}\right\rangle $
indicates how close we are to the regime $\left\langle \boldsymbol{p}_{\textrm{agree}}\right\rangle \approx0$
where the measurements strongly disturb each other for almost all
states, or the regime $\left\langle \boldsymbol{p}_{\textrm{agree}}\right\rangle \approx1$
where the mutual disturbance is inconsequential for almost all states.
We can therefore utilize $\left\langle \boldsymbol{p}_{\textrm{agree}}\right\rangle $
as a characteristic function that quantifies the relevance of the
uncertainty principle and outlines the transition between quantum
and classical regimes.

In order to calculate the value of $\left\langle \boldsymbol{p}_{\textrm{agree}}\right\rangle $
as a function of measurement resolution, we turn to the canonical
setting of finite-dimensional quantum mechanics. In this setting we
consider a particle on a periodic one-dimensional lattice with $d$
lattice sites. Initially, both lattice units of position and momentum
will be set to unity $\delta x\equiv1$, $\delta p\equiv1$. Later,
we will introduce proper units and consider the continuum limit.

Following the construction in \citep{vourdas2004quantum,jagannathan1981finite},
the Hilbert space of our system is given by the span of position basis
$\ket{X;n}$ for $n=0,...,d-1$. The momentum basis are related to
the position basis via the discrete Fourier transform $F$ 
\begin{align}
\ket{X;n} & =F^{\dagger}\ket{P;n}=\frac{1}{\sqrt{d}}\sum_{m=0}^{d-1}e^{-i2\pi mn/d}\ket{P;m}\\
\ket{P;m} & =F\ket{X;m}=\frac{1}{\sqrt{d}}\sum_{n=0}^{d-1}e^{i2\pi mn/d}\ket{X;n}.
\end{align}

\begin{figure}[t]
\includegraphics[width=1\columnwidth]{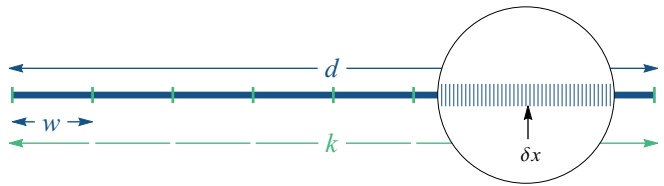}

\caption{\label{fig:lattice diagram}Periodic one dimensional lattice with
$d$ lattice sites in total, $w$ lattice sites in each coarse-graining
interval, and $k=d/w$ intervals. The lattice unit of length is $\delta x$.}
\end{figure}

In principle, realistic finite resolution measurements should be modeled
as unsharp POVMs \citep{busch1996quantum,peres2006quantum}, where
each POVM element is centered around a certain outcome value but has
a non-zero probability (usually Gaussian) to respond to the adjacent
values as well. In order to simplify the calculations we will consider
an idealized version of that in the form of coarse-grained projective
measurements. That is, each POVM element is a projection on a subspace
associated with a range of values such that an outcome associate with
each projection does not distinguish between any of the values in
the range.

We introduce the integer parameters $w_{x}$, $w_{p}$ to specify
the widths of the coarse-graining intervals for the corresponding
observables (larger $w$ means lower resolution). The variable $k=d/w$
specifies the number of coarse-graining intervals which we will also
assume to be an integer. See Fig. \ref{fig:lattice diagram} for a
diagrammatic summary of the relevant lengths.

The coarse-grained position and momentum observables are constructed
from the spectral projections
\begin{align*}
\Pi_{X;\nu} & =\sum_{n=\nu w_{x}}^{\nu w_{x}+w_{x}-1}\ket{X;n}\bra{X;n}\\
\Pi_{P;\mu} & =\sum_{m=\mu w_{p}}^{\mu w_{p}+w_{p}-1}\ket{P;m}\bra{P;m}
\end{align*}
associated with the eigenvalues of coarse-grained position $\nu=0,...,k_{x}-1$
and momentum $\mu=0,...,k_{p}-1$. The coarse-grained observables
are then given by 
\[
X_{cg}=\sum_{\nu=0}^{k_{x}-1}\nu\,\Pi_{X;\nu}\,\,\,\,\,\,\,\,\,\,\,\,\,\,\,P_{cg}=\sum_{\mu=0}^{k_{p}-1}\mu\,\Pi_{P;\mu}.
\]
In the following, we only compute the probabilities of outcomes so
$P_{cg}$ and $X_{cg}$ are only shown here for the sake of completeness;
the spectral projections $\Pi_{X;\nu}$ and $\Pi_{P;\mu}$ is all
we need.

Let us now calculate the probability of getting the outcomes $\nu,\mu,\nu$
in an instantaneous sequence of position-momentum-position measurements
on the initial state $\rho$. If $\rho^{\left(\nu\right)}$, $\rho^{\left(\nu\mu\right)}$
are the intermediate post-measurement states in this sequence then
we can express this probability as 
\begin{align}
\boldsymbol{p}_{xpx}\left(\nu,\mu,\nu|\rho\right) & =tr\left[\Pi_{X;\nu}\rho\right]tr\left[\Pi_{P;\mu}\rho^{\left(\nu\right)}\right]tr\left[\Pi_{X;\nu}\rho^{\left(\nu\mu\right)}\right]\nonumber \\
 & =tr\left[\left(\Pi_{X;\nu}\Pi_{P;\mu}\Pi_{X;\nu}\right)^{2}\rho\right]\label{eq:def p_xpx}
\end{align}
where the last line follows using explicit expressions for $\rho^{\left(\nu\right)}$
and $\rho^{\left(\nu\mu\right)}$. Then, the probability that both
position outcomes agree, regardless of the outcomes, is 
\begin{align}
\boldsymbol{p}_{\textrm{agree}}\left(\rho\right) & =\sum_{v=0}^{k_{x}-1}\sum_{\mu=0}^{k_{p}-1}\boldsymbol{p}_{xpx}\left(\nu,\mu,\nu|\rho\right)\nonumber \\
 & =tr\left[\sum_{v=0}^{k_{x}-1}\sum_{\mu=0}^{k_{p}-1}\left(\Pi_{X;\nu}\Pi_{P;\mu}\Pi_{X;\nu}\right)^{2}\rho\right].\label{eq:p_agree}
\end{align}
From Eq. (\ref{eq:p_agree}) we identify the observable 
\[
\Lambda_{\textrm{agree}}=\sum_{\nu=0}^{k_{x}-1}\sum_{\mu=0}^{k_{p}-1}\left(\Pi_{X;\nu}\Pi_{P;\mu}\Pi_{X;\nu}\right)^{2}
\]
whose expectation values are the probabilities $\boldsymbol{p}_{\textrm{agree}}\left(\rho\right)=tr\left(\Lambda_{\textrm{agree}}\rho\right)$.

Since $\boldsymbol{p}_{\textrm{agree}}\left(\rho\right)$ is linear
in $\rho$, the average $\left\langle \boldsymbol{p}_{\textrm{agree}}\right\rangle $
is given by $\boldsymbol{p}_{\textrm{agree}}\left(\left\langle \rho\right\rangle \right)$
where $\left\langle \rho\right\rangle =\frac{1}{d}I$ is the average
state. We can then calculate 
\begin{align}
\left\langle \boldsymbol{p}_{\textrm{agree}}\right\rangle  & =\boldsymbol{p}_{\textrm{agree}}\left(\frac{1}{d}I\right)=\frac{1}{d}tr\left[\Lambda_{\textrm{agree}}\right]\nonumber \\
 & =\frac{w_{x}}{d}+\frac{2}{w_{x}w_{p}d}\sum_{n=1}^{w_{p}-1}\left(w_{p}-n\right)\frac{\sin^{2}\left(\frac{\pi nw_{x}}{d}\right)}{\sin^{2}\left(\frac{\pi n}{d}\right)}\label{eq: lam_agree expr plane}
\end{align}
(see Appendix \ref{app: calc of p_agree} for the details of this
calculation).
\begin{figure}[t]
\begin{raggedright}
(a)
\par\end{raggedright}
\includegraphics[width=1\columnwidth]{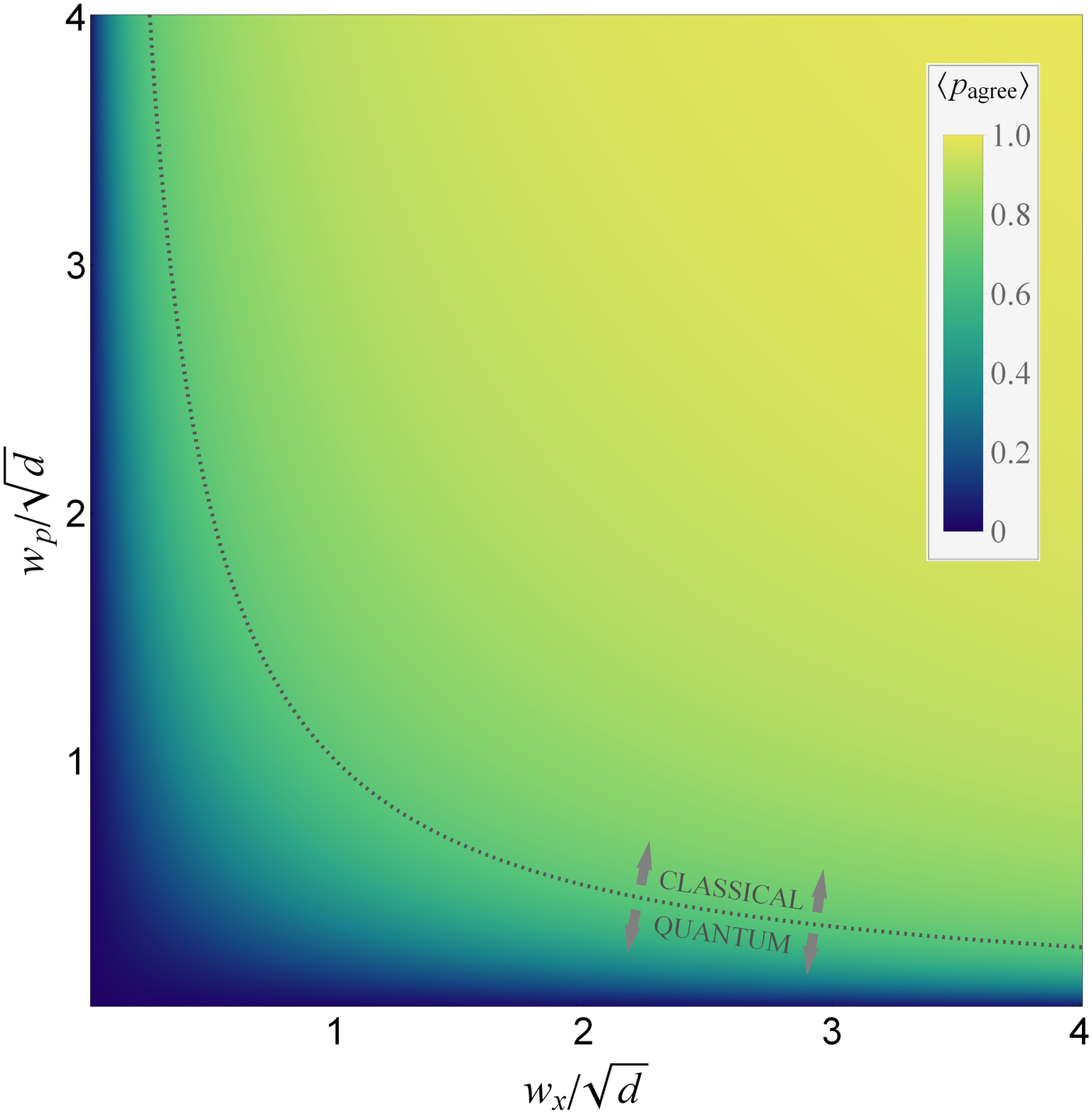}
\begin{raggedright}
(b)
\par\end{raggedright}
\includegraphics[width=1\columnwidth]{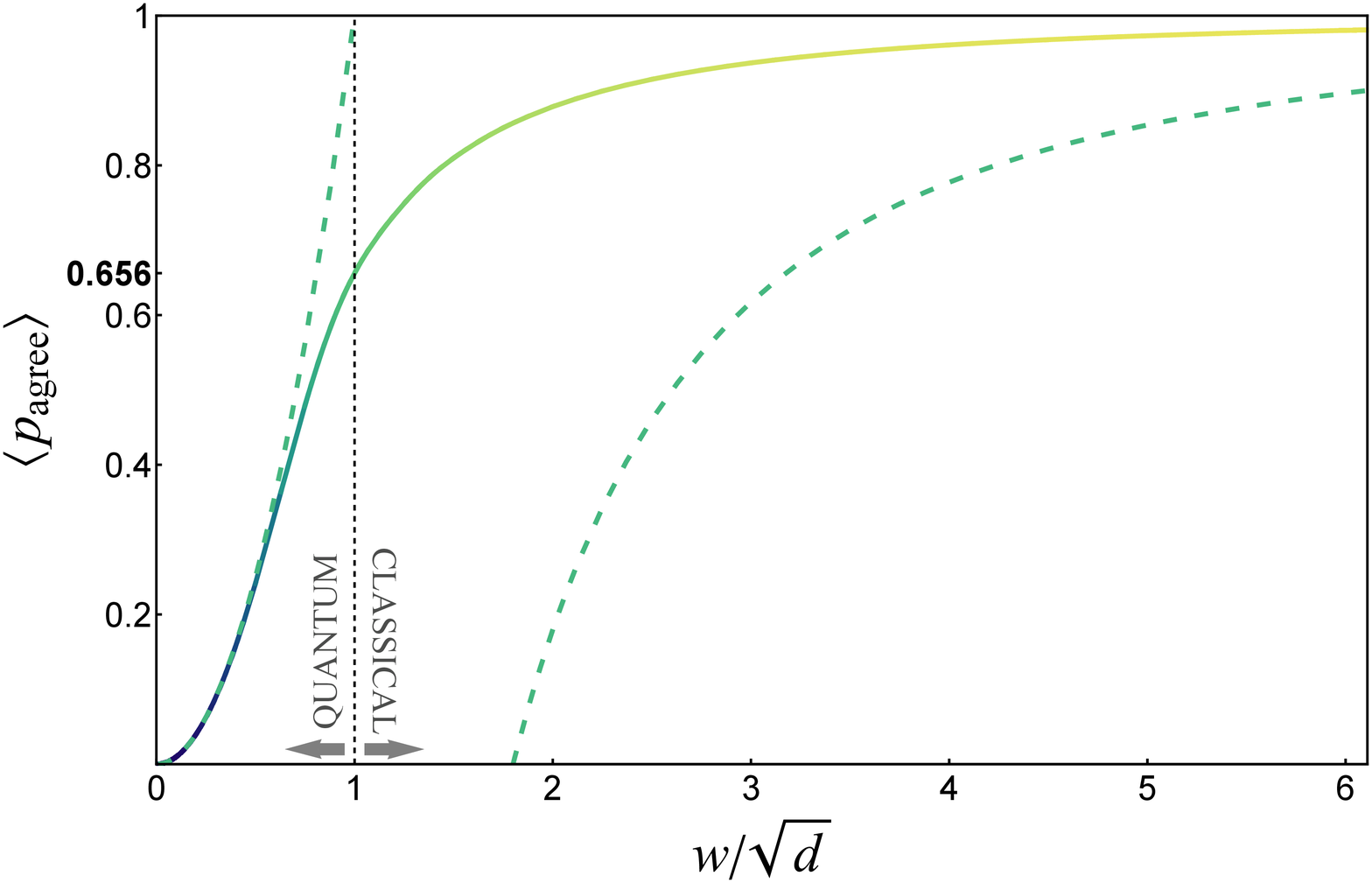}\caption{\label{fig:ClassicalityVw}(a) The plot of the average probability
$\left\langle \boldsymbol{p}_{\textrm{agree}}\right\rangle $ that
an instantaneous succession of position-momentum-position measurements
will agree on both outcomes of position as a function of the resolution
parameters $w_{x}$, $w_{p}$ on a lattice of length $d$. The dotted
curve $w_{x}w_{p}=d$ is the boundary that outlines the intermediate
scale with respect to which we distinguish the quantum and classical
regimes. (b) The plot of $\left\langle \boldsymbol{p}_{\textrm{agree}}\right\rangle $
(solid) on the diagonal $w=w_{x}=w_{p}$ with the upper and lower
bounds (dashed) from Eqs. (\ref{eq: lam agree upper bound}) and (\ref{eq:  lam agree lower bound}).}
\end{figure}

The plot of $\left\langle \boldsymbol{p}_{\textrm{agree}}\right\rangle $
as a function of $w_{x}$, $w_{p}$ is shown in Fig. \ref{fig:ClassicalityVw}(a)
which makes it clear that $\left\langle \boldsymbol{p}_{\textrm{agree}}\right\rangle $
is symmetric under the exchange of $w_{x}$ with $w_{p}$. The plot
of $\left\langle \boldsymbol{p}_{\textrm{agree}}\right\rangle $ along
the diagonal $w=w_{x}=w_{p}$ is shown in Fig. \ref{fig:ClassicalityVw}(b)
together with the upper and lower bounds
\begin{align}
 & \left\langle \boldsymbol{p}_{\textrm{agree}}\right\rangle \leq w^{2}/d &  & w<\sqrt{d}\label{eq: lam agree upper bound}\\
 & \left\langle \boldsymbol{p}_{\textrm{agree}}\right\rangle \geq1-\frac{2}{\pi^{2}}\frac{\ln\left(w^{2}/d\right)+3\pi^{2}/2}{w^{2}/d} &  & w>\sqrt{d}\label{eq:  lam agree lower bound}
\end{align}
(see Appendix \ref{app: calc of bounds} for the derivation). Note
that $\sqrt{d}$ distinguishes the two separate domains where these
bounds are valid.

The upper bound (\ref{eq: lam agree upper bound}) tells us that when
$w<\sqrt{d}$, the value of $\left\langle \boldsymbol{p}_{\textrm{agree}}\right\rangle $
falls to $0$ at least as fast as $\sim w^{2}$. The lower bound (\ref{eq:  lam agree lower bound})
tells us that when $w>\sqrt{d}$, the value of $\left\langle \boldsymbol{p}_{\textrm{agree}}\right\rangle $
climbs to $1$ at least as fast as $\sim1-\frac{\ln w^{2}}{w^{2}}$.
The fact that the domains of these bounds are separated by $\sqrt{d}$,
implies that there is an inflection in $\left\langle \boldsymbol{p}_{\textrm{agree}}\right\rangle $
somewhere around $w=\sqrt{d}$ along the diagonal $w=w_{x}=w_{p}$.
That is, $\sqrt{d}$ is an intermediate scale where neither bound
applies so it can serve as a reference point with respect to which
we distinguish the quantum and classical regimes.

The above observation can be extended to the entire plane of $w_{x}$,
$w_{p}$, where the curve $w_{x}w_{p}=d$ generalizes the point $w=\sqrt{d}$.
According to the plot in Fig. \ref{fig:ClassicalityVw}(a), as we
get farther from the curve $w_{x}w_{p}=d$, we get deeper into one
of the regimes, and an inflection in $\left\langle \boldsymbol{p}_{\textrm{agree}}\right\rangle $
occurs somewhere near the curve. It can be shown (see Appendix \ref{app: p_agree on the curve})
that the intermediate value $\left\langle \boldsymbol{p}_{\textrm{agree}}\right\rangle \approx0.656$
holds almost everywhere on this curve, except the far ends where it
climbs to $1$.

There is nothing special about the value $0.656$, however, the significance
of the curve $w_{x}w_{p}=d$ is that it outlines the intermediate
scale in phase space with respect to which we can distinguish the
quantum regime from the classical. That is, the curve $w_{x}w_{p}=d$
sets a reference scale so we can say that
\[
\begin{cases}
\left\langle \boldsymbol{p}_{\textrm{agree}}\right\rangle \approx1 & w_{x}w_{p}\gg d\\
\left\langle \boldsymbol{p}_{\textrm{agree}}\right\rangle \approx0 & w_{x}w_{p}\ll d.
\end{cases}
\]
We can of course say the same about $w_{x}w_{p}=cd$ for some $c\neq1$;
the important fact is that the constraint $cd$ depends linearly on
$d$. We will see below that for $c=1$ this constraint corresponds
exactly to the Planck constant.

\section{The continuum limit and the implications of minimal length}

\subsection{Perturbations of $\left\langle \boldsymbol{p}_{\textrm{agree}}\right\rangle $}

We will now introduce proper units. The total length of the lattice
in proper units is $L=\delta xd$, where $\delta x$ is the smallest
unit of length associated with one lattice spacing. The smallest unit
of inverse length, or a wavenumber, is then $1/L$. With the de Broglie
relation $p=2\pi\hbar/\lambda$, we can convert wavenumbers $1/\lambda$
to momenta, so the smallest unit of momentum is $\delta p=2\pi\hbar/L$.\footnote{Note that the de Broglie relation is the source of the Planck constant
in all of the following equations} The coarse-graining intervals $w_{x}$ and $w_{p}$ become $\Delta x=\delta xw_{x}$
and $\Delta p=\delta pw_{p}$ when expressed in proper units.

The continuum limit can be achieved by taking $\delta x\rightarrow0$
and $d\rightarrow\infty$ while keeping $L$ constant. The coarse-graining
interval of position $\Delta x=\delta xw_{x}$ is kept constant as
well by fixing the total number of intervals $k_{x}=d/w_{x}$ while
$w_{x}\rightarrow\infty$. Unlike $\delta x$, $\delta p=2\pi\hbar/L$
does not vanish in the continuum limit (the momentum of a particle
in a box remains quantized) so the coarse-graining intervals of momentum
$\Delta p=\delta pw_{p}$ are unaffected and $w_{p}$ remains a finite
integer.

We may now ask what happens to $\left\langle \boldsymbol{p}_{\textrm{agree}}\right\rangle $
as we take the continuum limit. Since $w_{x}/d=\Delta x/L$, the expression
in Eq. (\ref{eq: lam_agree expr plane}) can be re-expressed using
the proper units of length as 
\begin{equation}
\left\langle \boldsymbol{p}_{\textrm{agree}}\right\rangle =\frac{\Delta x}{L}+\frac{L}{\Delta x}\frac{2}{w_{p}}\sum_{n=1}^{w_{p}-1}\left(w_{p}-n\right)\frac{\sin^{2}\left(\frac{\pi n\Delta x}{L}\right)}{\left[d\sin\left(\frac{\pi n}{d}\right)\right]^{2}}.\label{eq: lam_agree expr plane cont}
\end{equation}
We did not have to use the proper units of momentum since 
\[
w_{p}=\frac{\Delta p}{\delta p}=\frac{\Delta p}{2\pi\hbar}L,
\]
which is a legitimate quantity even in the continuum limit (provided
that $L$ is finite).

The only evidence for the lattice structure that remains in Eq. (\ref{eq: lam_agree expr plane cont})
is the $d$-dependence of the factors 
\[
\left[d\sin\left(\frac{\pi n}{d}\right)\right]^{-2}=\frac{1}{\pi^{2}n^{2}}+\frac{1}{3d^{2}}+O\left(\frac{1}{d^{3}}\right).
\]
In the continuum limit they reduce to $1/\pi^{2}n^{2}$, but when
the minimal length $\delta x=L/d$ is above $0$, these factors are
perturbed with the leading order contribution of $1/3d^{2}$.

\begin{figure}[t]
\begin{raggedright}
(a)
\par\end{raggedright}
\includegraphics[width=1\columnwidth]{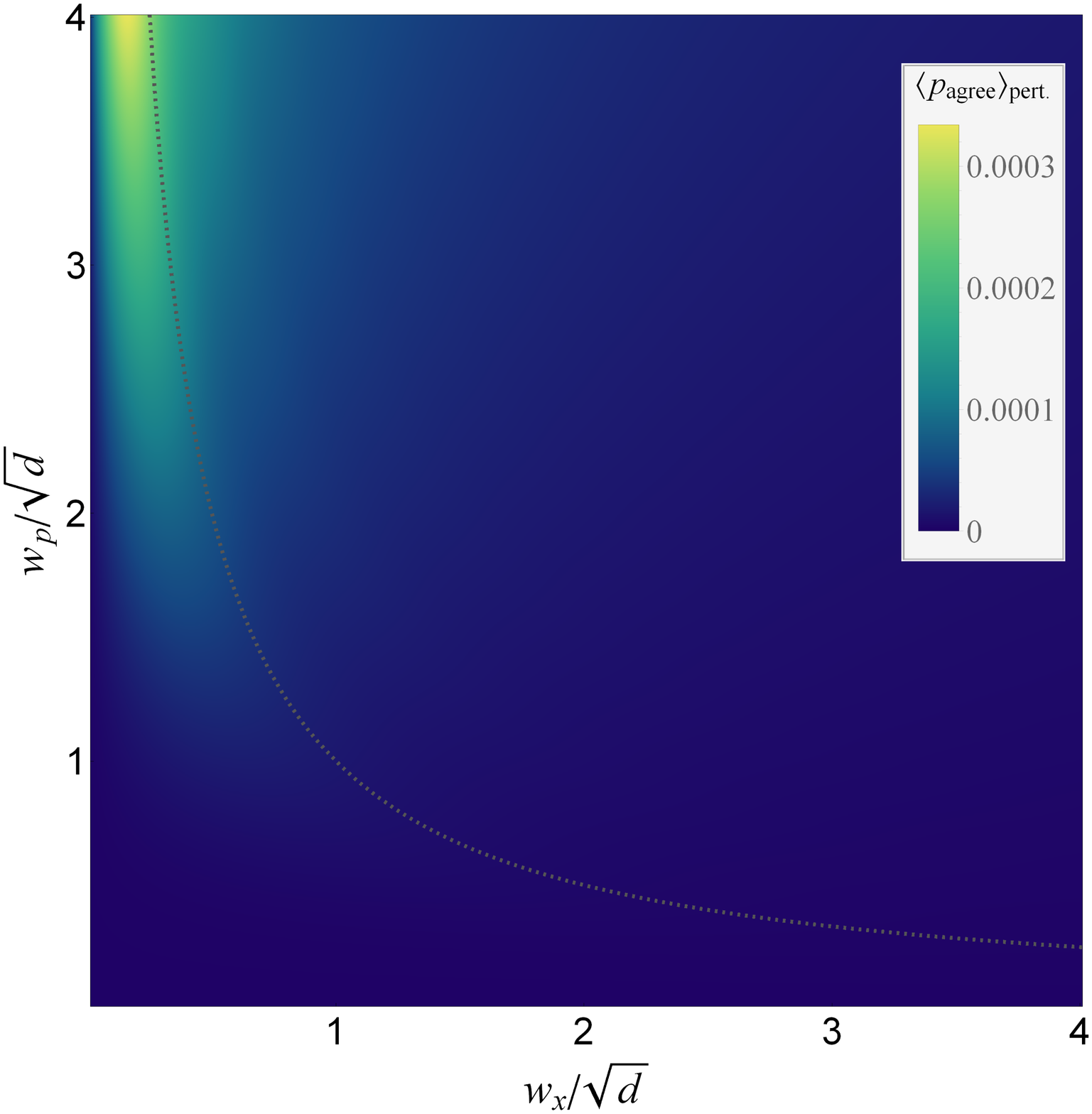}
\begin{raggedright}
(b)
\par\end{raggedright}
\includegraphics[width=1\columnwidth]{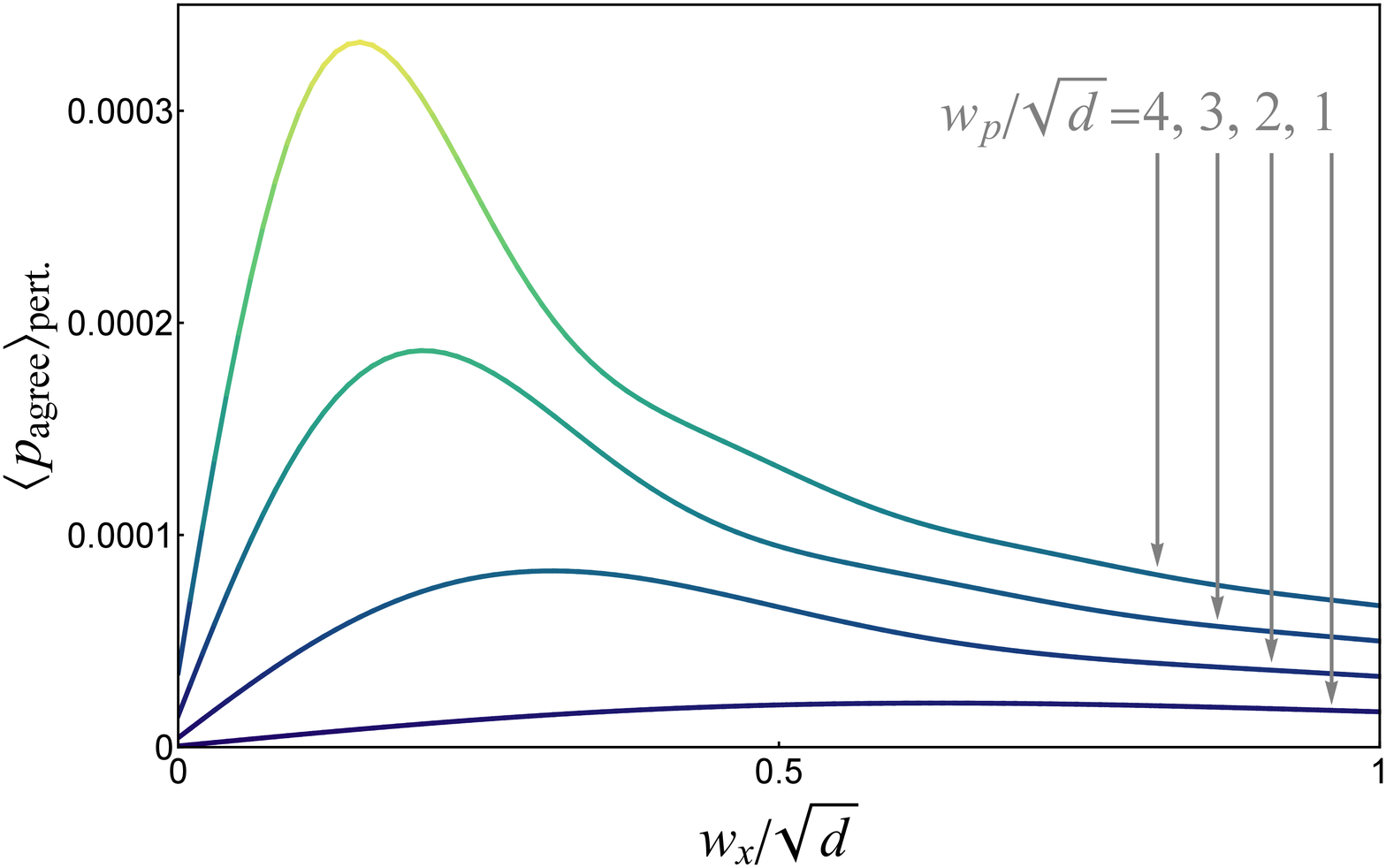}

\caption{\label{fig:perturbation}(a) The plot of the perturbation of $\left\langle \boldsymbol{p}_{\textrm{agree}}\right\rangle $
on a lattice of length $d=10^{4}$ as given by Eq. (\ref{eq:p_agree_pert_ex})
(the dotted curve is $w_{x}w_{p}=d$). (b) The profile of the perturbation
as given by Eq. (\ref{eq:p_agree_pert}) for $w_{p}/\sqrt{d}=1,2,3,4$
and $d=10^{4}$.}
\end{figure}

The leading order perturbation term of $\left\langle \boldsymbol{p}_{\textrm{agree}}\right\rangle $
is therefore
\begin{align}
\left\langle \boldsymbol{p}_{\textrm{agree}}\right\rangle _{\textrm{pert.}} & =\frac{L}{\Delta x}\frac{2}{w_{p}}\sum_{n=1}^{w_{p}-1}\left(w_{p}-n\right)\frac{\sin^{2}\left(\frac{\pi n\Delta x}{L}\right)}{3d^{2}}\nonumber \\
 & =\frac{2}{3w_{x}w_{p}d}\sum_{n=1}^{w_{p}-1}\left(w_{p}-n\right)\sin^{2}\left(\frac{\pi nw_{x}}{d}\right)\label{eq:p_agree_pert_ex}
\end{align}
and we reverted to the lattice units in the last step. In Fig. \ref{fig:perturbation}(a)
we have plotted Eq. (\ref{eq:p_agree_pert_ex}) for $d=10^{4}$. As
we can see from the plot, the lattice perturbation gets stronger as
$w_{x}$ decreases and $w_{p}$ increases, and the perturbation spikes
in the regime where $w_{x}<\sqrt{d}$ and $w_{p}>\sqrt{d}$.

Focusing on this regime, we can assume that $w_{p}\gg1$ (since $\sqrt{d}\gg1$)
and approximate the sum with an integral. That is,
\begin{align}
\left\langle \boldsymbol{p}_{\textrm{agree}}\right\rangle _{\textrm{pert.}} & =\frac{2w_{p}}{3w_{x}d}\sum_{n=1}^{w_{p}-1}\frac{1}{w_{p}}\left(1-\frac{n}{w_{p}}\right)\sin^{2}\left(\frac{\pi nw_{x}w_{p}}{w_{p}d}\right)\nonumber \\
 & \approx\frac{2}{3}\frac{w_{p}}{w_{x}d}\int_{0}^{1}d\alpha\left(1-\alpha\right)\sin^{2}\left(\pi\alpha\frac{w_{x}w_{p}}{d}\right)\nonumber \\
 & =\frac{2}{3}\frac{w_{p}}{w_{x}d}\left[\frac{1}{4}+\frac{\cos(2\pi w_{x}w_{p}/d)-1}{8\pi^{2}\left(w_{x}w_{p}/d\right)^{2}}\right].\label{eq:p_agree_pert}
\end{align}
See Fig. \ref{fig:perturbation}(b) for the plot of Eq. (\ref{eq:p_agree_pert}).
By re-introducing proper units and rearranging we get
\[
\left\langle \boldsymbol{p}_{\textrm{agree}}\right\rangle _{\textrm{pert.}}\approx\frac{1}{6\pi}\left(\frac{\delta x}{\Delta x}\right)^{2}\left[\frac{1}{2}\frac{\Delta x\Delta p}{\hbar}+\frac{\cos(\Delta x\Delta p/\hbar)-1}{\Delta x\Delta p/\hbar}\right].
\]
In particular, on the curve $\Delta x\Delta p=2\pi\hbar$ we have
$\left\langle \boldsymbol{p}_{\textrm{agree}}\right\rangle _{\textrm{pert.}}\approx\frac{1}{6}\left(\delta x/\Delta x\right)^{2}$
so the perturbation keeps growing as we ascend on this curve.

Since $\left\langle \boldsymbol{p}_{\textrm{agree}}\right\rangle $
is an operationally defined quantity, it in principle can be measured.
The perturbation term $\left\langle \boldsymbol{p}_{\textrm{agree}}\right\rangle _{\textrm{pert.}}$
can therefore be leveraged as a signal of the discontinuity of space
in experimental approaches. That is, given the continuum probabilities
\[
\left\langle \boldsymbol{p}_{\textrm{agree}}\right\rangle _{\textrm{cont.}}=\frac{\Delta x}{L}+\frac{L}{\Delta x}\frac{2}{w_{p}}\sum_{n=1}^{w_{p}-1}\left(w_{p}-n\right)\frac{\sin^{2}\left(\frac{\pi n\Delta x}{L}\right)}{\pi^{2}n^{2}}
\]
we expect to find that 
\[
\left\langle \boldsymbol{p}_{\textrm{agree}}\right\rangle =\left\langle \boldsymbol{p}_{\textrm{agree}}\right\rangle _{\textrm{cont.}}+\left\langle \boldsymbol{p}_{\textrm{agree}}\right\rangle _{\textrm{pert.}}+O\left(\frac{1}{d^{3}}\right)
\]
so by measuring the deviation of $\left\langle \boldsymbol{p}_{\textrm{agree}}\right\rangle $
from the value of $\left\langle \boldsymbol{p}_{\textrm{agree}}\right\rangle _{\textrm{cont.}}$
as defined above, we can detect the discontinuity of space.

For realistic values of $d$ the signal of $\left\langle \boldsymbol{p}_{\textrm{agree}}\right\rangle _{\textrm{pert.}}$
is of course extremely weak. However, the ``humps'' of $\left\langle \boldsymbol{p}_{\textrm{agree}}\right\rangle _{\textrm{pert.}}$
start to appear on the intermediate scales of $\Delta x<\delta x\sqrt{d}$
and $\Delta p>\delta p\sqrt{d}$ (see Fig. \ref{fig:perturbation}(b)),
so we do not have to go to the extremes of minimal length or maximal
momentum to look for them.

\subsection{Factorizing the Planck constant}

With the introduction of proper units we observe that the smallest
unit of phase space area on a lattice is\footnote{This is a well known constraint that comes up in the construction
of Generalized Clifford Algebras in finite-dimensional quantum mechanics.
See \citep{singh2018modeling} for an overview and the references
therein.} $\delta x\delta p=2\pi\hbar/d$. Therefore, the curve $w_{x}w_{p}=d$
that outlines the intermediate scale in phase space becomes 
\begin{equation}
\Delta x\Delta p=\delta x\delta p\,w_{x}w_{p}=\delta x\delta p\,d=2\pi\hbar.\label{eq:class. boundary with units}
\end{equation}
Thus, we have recovered Heisenberg's original argument that the Planck
constant sets the scale in phase space where the mutual disturbance
effects become significant. Note that Eq. (\ref{eq:class. boundary with units})
is related to what is known as the \emph{error-disturbance uncertainty
relations }(not to be confused with the \emph{preparation uncertainty
relations})\emph{.} We thus see that in the unitless lattice setting
(where $\delta x\equiv1$ and $\delta p\equiv1$) the constant $d$
is the unitless ``Planck constant''. \footnote{Note that unlike $2\pi\hbar$, the constant $d$ depends on the size
of the system. This inconstancy traces back to the fact that in the
unitless case we define $\delta p\equiv1$, while in proper units
we have $\delta p=2\pi\hbar/L$, which depends on the total length
$L$.}

In the continuous phase space, the uncertainty principle is only associated
with the constant $2\pi\hbar$, which does not admit a preferred factorization
into position and momentum. On the lattice, however, the same constant
is given by $\delta x\delta pd$, which can be factorized as $\delta x\sqrt{d}$
and $\delta p\sqrt{d}$. This factorization is not arbitrary and the
significance of the scales $\delta x\sqrt{d}$ and $\delta p\sqrt{d}$
is supported by the analysis of $\left\langle \boldsymbol{p}_{\textrm{agree}}\right\rangle $.
In particular, we saw that the perturbation $\left\langle \boldsymbol{p}_{\textrm{agree}}\right\rangle _{\textrm{pert.}}$
due to the discontinuity of the lattice spikes in the regime where
$\Delta x<\delta x\sqrt{d}$ and $\Delta p>\delta p\sqrt{d}$.

The significance of the scale $\sqrt{d}$ on a lattice can also be
observed from $\left\langle \boldsymbol{p}_{\textrm{agree}}\right\rangle $
directly. In Fig. \ref{fig:ClassicalityVw}(a) we can see that when
the localization in position $w_{x}$ crosses $\sqrt{d}$ from above,
the localization in momentum $w_{p}$ has to diverge faster than it
converges in $w_{x}$ in order to stay in the classical regime. In
contrast, as long as both $w_{x},w_{p}\gg\sqrt{d}$, the classical
regime is insensitive to the variations in these variables and there
is no need to compensate the increase in localization for one variable
with the decrease in localization for the other.

This observation is directly analogous to the analysis of Kofler and
Brukner in \citep{kofler2007classical} (similar question have been
considered in \citep{poulin2005macroscopic} and \citep{peres2006quantum})
where they have demonstrated that for a spin-$j$ system, incompatible
spin components can simultaneously be determined if the resolution
of measurements is coarse compared to $\sqrt{j}$. Our analysis show
that the same conclusion applies to position and momentum on a lattice,
where both variables can simultaneously be determined if the resolution
of measurements is coarse compared to $\sqrt{d}$.

The uncertainty principle on a lattice can therefore primarily be
associated with the unitless scale $\sqrt{d}$, which identifies the
intermediate scales $\delta x\sqrt{d}$ and $\delta p\sqrt{d}$ for
position and momentum. The intermediate scale in phase space is in
turn given by

\[
\left(\delta x\sqrt{d}\right)\left(\delta p\sqrt{d}\right)=\delta x\delta p\,d=2\pi\hbar.
\]

The intermediate length scale $\delta x\sqrt{d}$ can be identified
as the scale around which increases in localization in position result
in\emph{ equal} decreases in localization in momentum, and vice versa.
Of course, this definition is only meaningful on a lattice because
it requires the fundamental units $\delta x$ and $\delta p$ in terms
of which we can compare the changes in localization for both variables.
Nevertheless, we conclude that on a lattice, in addition to the minimal
length $\delta x$ and the maximal length $L$, the uncertainty principle
singles out another significant length 
\[
l_{u}=\delta x\sqrt{d}.
\]

The length $l_{u}$ is directly related to the minimal length $\delta x$
via $L=\delta xd$ as $l_{u}=\sqrt{\delta x\,L}$ or $\delta x=l_{u}^{2}/L$.
The length $l_{u}$ is therefore the geometric mean of the minimal
length $\delta x$ and the maximal length $L$. It can also be framed
as the length for which there are as many intervals $l_{u}$ in $L$
as there are $\delta x$ in $l_{u}$. In the continuum limit, where
the minimal length $\delta x$ vanishes, the length $l_{u}=\sqrt{\delta x\,L}$
must also vanish. Therefore, if we can establish that $l_{u}>0$ then
it follows that $\delta x>0$.

We saw that the perturbations of $\left\langle \boldsymbol{p}_{\textrm{agree}}\right\rangle $
spike in the regime where $\Delta x<l_{u}$, but it is not clear at
this point what realistically observable effects can be associated
with the length $l_{u}$. If such effects can be identified, however,
then the discontinuity of space can be probed at scales that are many
orders of magnitude greater than the Planck length. For instance,
for $L\sim1\,\textrm{m}$ of the order of a macroscopic box and $\delta x\sim10^{-35}\,\textrm{m}$
of the order of Planck length, we have $l_{u}\sim10^{-17.5}\,\textrm{m}$
which is much closer to the scale of experiments.

\section{Conclusion}

In the present work we have studied the effects of the uncertainty
principle on a finite-dimensional periodic lattice, and their dependence
on minimal length. Instead of modifying the canonical commutation
relations, we have operationally quantified the mutual disturbance
effects with the average probability $\left\langle \boldsymbol{p}_{\textrm{agree}}\right\rangle $,
and compared it to the continuum limit.

The analysis of $\left\langle \boldsymbol{p}_{\textrm{agree}}\right\rangle $
indicated that $\sqrt{d}$ is a significant scale on a lattice that
separates the classical regime of joint measurability, from the quantum
regime where mutual disturbance effect are important. In the units
of length, the scale $\sqrt{d}$ corresponds to the geometric mean
$l_{u}=\sqrt{\delta xL}$ of the minimal length $\delta x$ and the
maximal length $L$, and in phase space it corresponds to the Planck
constant. This result is consistent with the conclusion of Kofler
and Brukner \citep{kofler2007classical} for spin-$j$ systems where
incompatible observables can simultaneously be determined if the resolution
of measurements is coarse compared to $\sqrt{j}$.

We have also analyzed the perturbations of $\left\langle \boldsymbol{p}_{\textrm{agree}}\right\rangle $
due to the non-vanishing minimal length $\delta x$ on a lattice.
As a result, we saw that the perturbations become pronounced in the
regime where the resolution in position falls below the scale of $l_{u}$,
and the resolution in momentum rises above the scale of $\delta p\sqrt{d}$.

This is a preliminary result and we make no attempt to translate it
into experimental predictions. For a more concrete experimental proposal
it will be necessary to repeat the analysis of $\boldsymbol{p}_{\textrm{agree}}$
with the experimentally accessible ensemble of states $\rho$. Furthermore,
depending on the experimental implementation, it will be necessary
to use the non-idealized coarse-grained measurements and (possibly)
account for the time evolution inbetween or during the measurements.
Nonetheless, this result indicates that in principle it is possible
to detect the discontinuity of the underlying space on the intermediate
scales associated with $\sqrt{d}$.
\begin{acknowledgments}
The author would like to thank Ashmeet Singh, Jason Pollack, Pedro
Lopes, Michael Zurel, \v{C}aslav Brukner and Robert Raussendorf for
helpful comments and discussions, and Rita Livshits for proofreading
the manuscript. Special thanks to the anonymous referee whose feedback
helped elucidate some of the arguments. This work was supported by
the Natural Sciences and Engineering Research Council of Canada (NSERC).
\end{acknowledgments}

\appendix
\onecolumngrid

\section{{\normalsize{}General definitions and identities}}

As described above, we are dealing with the $d$-dimensional Hilbert
space of a particle on a periodic lattice with the position and momentum
basis related via the discrete Fourier transform $F$. The translation
operators $T_{X}$, $\text{\ensuremath{T_{P}} }$ in position and
momentum can be defined by their action on the basis \citep{vourdas2004quantum}
as follows 
\begin{align*}
T_{X}\ket{X;n} & =\ket{X;n+1} &  & T_{X}^{\dagger}\ket{X;n}=\ket{X;n-1}\\
T_{P}\ket{P;m} & =\ket{P;m+1} &  & T_{P}^{\dagger}\ket{P;m}=\ket{P;m-1}
\end{align*}
where $\pm1$ are $\mathsf{mod}\,d$. By expanding the position basis
in momentum basis and vice versa and using the definitions, it is
straight forward to verify that
\begin{align*}
T_{P}\ket{X;n} & =e^{i2\pi n/d}\ket{X;n} &  & T_{P}^{\dagger}\ket{X;n}=e^{-i2\pi n/d}\ket{X;n}\\
T_{X}\ket{P;m} & =e^{-i2\pi m/d}\ket{P;m} &  & T_{X}^{\dagger}\ket{P;m}=e^{i2\pi m/d}\ket{P;m}.
\end{align*}
Therefore, $T_{P}$ commutes with $\ket{X;n}\bra{X;n}$ and so does
$T_{X}$ with $\ket{P;m}\bra{P;m}$. This also means that $T_{P}$
commutes with $\Pi_{X;\nu}$ and $T_{X}$ commutes with $\Pi_{P;\mu}$.

Using the translation operators we can express the coarse-grained
position and momentum projections as 
\begin{align*}
\Pi_{X;\nu} & =T_{X}^{\nu w_{x}}\Pi_{X;0}T_{X}^{\nu w_{x}\dagger} &  & \Pi_{P;\mu}=T_{P}^{\mu w_{p}}\Pi_{P;0}T_{P}^{\mu w_{p}\dagger}.
\end{align*}
Then, using the commutativity of projections with translations we
get the identity

\begin{equation}
\Pi_{X;\nu}\Pi_{P;\mu}\Pi_{X;\nu}=T_{P}^{\mu w_{p}}\left(\Pi_{X;\nu}\Pi_{P;0}\Pi_{X;\nu}\right)T_{P}^{\mu w_{p}\dagger}=T_{P}^{\mu w_{p}}T_{X}^{\nu w_{x}}\left(\Pi_{X;0}\Pi_{P;0}\Pi_{X;0}\right)T_{X}^{\nu w_{x}\dagger}T_{P}^{\mu w_{p}\dagger}.\label{eq:PxPpPx=00003DP0P0P0}
\end{equation}

Focusing on the $\nu=\mu=0$ case we can express

\begin{equation}
\Pi_{X;0}\Pi_{P;0}\Pi_{X;0}=\sum_{m=0}^{w_{p}-1}\Pi_{X;0}\ket{P;m}\bra{P;m}\Pi_{X;0}=\frac{1}{k_{x}}\sum_{m=0}^{w_{p}-1}\ket{P_{0};m}\bra{P_{0};m}.\label{eq: P0P0P0 in terms of |P_0,m>}
\end{equation}
Thus, we define the truncated momentum states which are given by the
normalized support of the $m$'th momentum state on the $\nu$'th
position interval: 
\begin{align}
\ket{P_{\nu};m} & :=\sqrt{k_{x}}\,\Pi_{X;\nu}\ket{P;m}=\frac{1}{\sqrt{w_{x}}}\sum_{n=\nu w_{x}}^{\nu w_{x}+w_{x}-1}e^{i2\pi mn/d}\ket{X;n}.\label{eq:def of P_nu}
\end{align}
In general, these states are not orthogonal and their overlap is given
by
\begin{align*}
\braket{P_{\nu'};m'}{P_{\nu};m} & =\delta_{\nu',\nu}k_{x}\bra{P;m'}\Pi_{X;\nu}\ket{P;m}=\delta_{\nu',\nu}\frac{k_{x}}{d}\sum_{n=\nu w_{x}}^{\nu w_{x}+w_{x}-1}e^{i2\pi\left(m-m'\right)n/d}.
\end{align*}

It will be convenient to identify sums such as the one above, by defining
the function 
\begin{equation}
\varDelta_{q}\left(x\right):=\frac{1}{q}\sum_{n=0}^{q-1}e^{i2\pi xn/q}=\frac{e^{i\pi\left(x-x/q\right)}}{q}\frac{\sin\left(\pi x\right)}{\sin\left(\pi x/q\right)}\label{eq:def of delta}
\end{equation}
over real $x$ and integer $q\geq1$. Note that $\varDelta_{q}\left(0\right)=1$.
Then, for $\nu'=\nu=0$ the overlap of truncated momentum states can
be expressed as 
\begin{align}
\braket{P_{0};m'}{P_{0};m} & =\varDelta_{w_{x}}\left(\frac{m-m'}{k_{x}}\right).\label{eq:P_0 inner prod}
\end{align}

\section{{\normalsize{}Calculation of Eq. (\ref{eq: lam_agree expr plane})\label{app: calc of p_agree}}}

Given the operator 
\[
\Lambda_{\textrm{agree}}=\sum_{\nu=0}^{k_{x}-1}\sum_{\mu=0}^{k_{p}-1}\left(\Pi_{X;\nu}\Pi_{P;\mu}\Pi_{X;\nu}\right)^{2}
\]
we are interested in the quantity $\left\langle \boldsymbol{p}_{\textrm{agree}}\right\rangle =\frac{1}{d}tr\left[\Lambda_{\textrm{agree}}\right]$.
Using the identity (\ref{eq:PxPpPx=00003DP0P0P0}) we can simplify
the problem:
\begin{equation}
\left\langle \boldsymbol{p}_{\textrm{agree}}\right\rangle =\frac{1}{d}tr\left[\sum_{\nu=0}^{k_{x}-1}\sum_{\mu=0}^{k_{p}-1}\left(\Pi_{X;\nu}\Pi_{P;\mu}\Pi_{X;\nu}\right)^{2}\right]=\frac{k_{x}k_{p}}{d}tr\left[\left(\Pi_{X;0}\Pi_{P;0}\Pi_{X;0}\right)^{2}\right].\label{eq: lam_agree in terms of P0P0P0}
\end{equation}

Using (\ref{eq: P0P0P0 in terms of |P_0,m>}) and (\ref{eq:P_0 inner prod})
we can further simplify 
\[
\left\langle \boldsymbol{p}_{\textrm{agree}}\right\rangle =\frac{1}{d}\frac{k_{p}}{k_{x}}\sum_{m,m'=0}^{w_{p}-1}\left|\braket{P_{0};m'}{P_{0};m}\right|^{2}=\frac{1}{d}\frac{k_{p}}{k_{x}}\sum_{m,m'=0}^{w_{p}-1}\left|\varDelta_{w_{x}}\left(\frac{m-m'}{k_{x}}\right)\right|^{2}.
\]
Since the summand depends only on the difference $n=m-m'$, we can
re-express the sum in terms of the single variable $n$
\[
\left\langle \boldsymbol{p}_{\textrm{agree}}\right\rangle =\frac{1}{d}\frac{k_{p}}{k_{x}}\sum_{n=-w_{p}+1}^{w_{p}-1}\left(w_{p}-\left|n\right|\right)\left|\varDelta_{w_{x}}\left(\frac{n}{k_{x}}\right)\right|^{2}.
\]
Since the summed function is symmetric $\left|\varDelta_{w_{x}}\left(x\right)\right|^{2}=\left|\varDelta_{w_{x}}\left(-x\right)\right|^{2}$,
we have
\[
\left\langle \boldsymbol{p}_{\textrm{agree}}\right\rangle =\frac{1}{d}\frac{k_{p}}{k_{x}}\left[w_{p}\left|\varDelta_{w_{x}}\left(0\right)\right|^{2}+2\sum_{n=1}^{w_{p}-1}\left(w_{p}-n\right)\left|\varDelta_{w_{x}}\left(\frac{n}{k_{x}}\right)\right|^{2}\right].
\]

Substituting the definition (\ref{eq:def of delta}) of $\varDelta_{w_{x}}$
and recalling that $\varDelta_{w_{x}}\left(0\right)=1$ and that $k_{x}=d/w_{x}$
and $k_{p}=d/w_{p}$, we get the result
\begin{align}
\left\langle \boldsymbol{p}_{\textrm{agree}}\right\rangle  & =\frac{1}{d}\frac{w_{x}}{w_{p}}\left[w_{p}+2\sum_{n=1}^{w_{p}-1}\left(w_{p}-n\right)\frac{1}{w_{x}^{2}}\frac{\sin^{2}\left(\frac{\pi nw_{x}}{d}\right)}{\sin^{2}\left(\frac{\pi n}{d}\right)}\right]\nonumber \\
 & =\frac{w_{x}}{d}+\frac{2}{w_{x}w_{p}d}\sum_{n=1}^{w_{p}-1}\left(w_{p}-n\right)\frac{\sin^{2}\left(\frac{\pi nw_{x}}{d}\right)}{\sin^{2}\left(\frac{\pi n}{d}\right)}.\label{eq: lam_agree in terms of wx wp}
\end{align}

The apparent asymmetry under the exchange of $w_{x}$ with $w_{p}$
in the result (\ref{eq: lam_agree in terms of wx wp}), traces back
to the apparent asymmetry under the exchange between $\Pi_{X;0}$
and $\Pi_{P;0}$ in the expression (\ref{eq: lam_agree in terms of P0P0P0}).
These asymmetries are only apparent because 
\[
tr\left[\left(\Pi_{X;0}\Pi_{P;0}\Pi_{X;0}\right)^{2}\right]=tr\left[\Pi_{X;0}\Pi_{P;0}\Pi_{X;0}\Pi_{P;0}\right]=tr\left[\left(\Pi_{P;0}\Pi_{X;0}\Pi_{P;0}\right)^{2}\right],
\]
so if we were to change the order in expression (\ref{eq: lam_agree in terms of P0P0P0})
to $tr\left[\left(\Pi_{P;0}\Pi_{X;0}\Pi_{P;0}\right)^{2}\right]$,
we would end up with
\[
\left\langle \boldsymbol{p}_{\textrm{agree}}\right\rangle =\frac{w_{p}}{d}+\frac{2}{w_{x}w_{p}d}\sum_{n=1}^{w_{x}-1}\left(w_{x}-n\right)\frac{\sin^{2}\left(\frac{\pi nw_{p}}{d}\right)}{\sin^{2}\left(\frac{\pi n}{d}\right)}.
\]
The form (\ref{eq: lam_agree in terms of wx wp}) is better suited
for the continuum limit where $w_{p}$ remains finite while $w_{x}$
is not (but $\frac{w_{x}}{d}$ is).

\section{{\normalsize{}The value of $\left\langle \boldsymbol{p}_{\textrm{agree}}\right\rangle $
on the curve $w_{x}w_{p}=d$\label{app: p_agree on the curve}}}

When $w_{x}w_{p}=d$ we can simplify
\begin{equation}
\left\langle \boldsymbol{p}_{\textrm{agree}}\right\rangle =\frac{1}{w_{p}}+\frac{2}{d^{2}}\sum_{n=1}^{w_{p}-1}\left(w_{p}-n\right)\frac{\sin^{2}\left(\frac{\pi n}{w_{p}}\right)}{\sin^{2}\left(\frac{\pi n}{d}\right)}.\label{eq: lam_agree on wxwp=00003Dd}
\end{equation}
First, let us consider the intermediate range of values $1\ll w_{p}\ll d$,
which includes $w_{p}=\sqrt{d}$ provided that $1\ll d$. Since $n<w_{p}\ll d$
we can approximate $\sin^{-2}\left(\frac{\pi n}{d}\right)\approx\left(\frac{\pi n}{d}\right)^{-2}$
and so
\begin{equation}
\left\langle \boldsymbol{p}_{\textrm{agree}}\right\rangle \approx\frac{1}{w_{p}}+\frac{2}{\pi^{2}}\sum_{n=1}^{w_{p}-1}\left(w_{p}-n\right)\frac{\sin^{2}\left(\frac{\pi n}{w_{p}}\right)}{n^{2}}.\label{eq: lam_agree on wxwp=00003Dd wp<<d}
\end{equation}
Since $1\ll w_{p}$, we can approximate the sum with an integral by
introducing the variable $\alpha=\frac{n}{w_{p}}\in\left[0,1\right]$
and $d\alpha=\frac{1}{w_{p}}$, such that
\begin{align*}
\left\langle \boldsymbol{p}_{\textrm{agree}}\right\rangle  & \approx\frac{1}{w_{p}}+\frac{2}{\pi^{2}}\sum_{n=1}^{w_{p}-1}\frac{1}{w_{p}}\left(1-\frac{n}{w_{p}}\right)\frac{\sin^{2}\left(\pi\frac{n}{w_{p}}\right)}{n^{2}/w_{p}^{2}}\\
 & \approx d\alpha+\frac{2}{\pi^{2}}\int_{0}^{1}d\alpha\left(1-\alpha\right)\frac{\sin^{2}\left(\pi\alpha\right)}{\alpha^{2}}\approx0.656.
\end{align*}
Thus, $\left\langle \boldsymbol{p}_{\textrm{agree}}\right\rangle \approx0.656$
for $1\ll w_{p}\ll d$ on the curve $w_{x}w_{p}=d$.

When the values of $w_{p}$ are close to $1$, we cannot assume that
$1\ll w_{p}$ but $w_{p}\ll d$ still holds so we can still use the
approximation (\ref{eq: lam_agree on wxwp=00003Dd wp<<d}). For $w_{p}=1$
the sum in (\ref{eq: lam_agree on wxwp=00003Dd wp<<d}) vanishes and
we are left with $\left\langle \boldsymbol{p}_{\textrm{agree}}\right\rangle =1/w_{p}=1$
(we can also see that from Eq. (\ref{eq: lam_agree in terms of P0P0P0})
that is easy to evaluate for $w_{p}=1$, and $w_{x}=d$). Numerically
evaluating Eq. (\ref{eq: lam_agree on wxwp=00003Dd wp<<d}) for the
subsequent values of $w_{p}$ results in the following series (considering
only $3$ significant figures)
\begin{center}
\begin{tabular}{|c|cccccccc|}
\hline 
$w_{p}$ & 1 & 2 & 3 & 4 & ... & 15 & 16 & ...\tabularnewline
$\left\langle \boldsymbol{p}_{\textrm{agree}}\right\rangle $ & $1.00$ & $0.703$ & $0.675$ & $0.667$ & ... & $0.657$ & $0.656$ & $0.656$\tabularnewline
\hline 
\end{tabular}
\par\end{center}

\noindent Thus, we can see that on one end of the curve $w_{x}w_{p}=d$,
where the $w_{p}$'s are small, the function $\left\langle \boldsymbol{p}_{\textrm{agree}}\right\rangle $
reaches and stays on the value $0.656$ starting from $w_{p}\geq16$.

Since $w_{x}$ and $w_{p}$ are interchangeable, we can re-express
Eq. (\ref{eq: lam_agree on wxwp=00003Dd wp<<d}) as 
\[
\left\langle \boldsymbol{p}_{\textrm{agree}}\right\rangle \approx\frac{1}{w_{x}}+\frac{2}{\pi^{2}}\sum_{n=1}^{w_{x}-1}\left(w_{x}-n\right)\frac{\sin^{2}\left(\frac{\pi n}{w_{x}}\right)}{n^{2}}.
\]
Then, on the other end of this curve, where the $w_{x}$'s are small,
the function $\left\langle \boldsymbol{p}_{\textrm{agree}}\right\rangle $
reaches and stays on the value $0.656$ starting from $w_{x}\geq16$.
Therefore, $\left\langle \boldsymbol{p}_{\textrm{agree}}\right\rangle \approx0.656$
almost everywhere on the curve $w_{x}w_{p}=d$, with the exception
of the far ends where $w_{p}<16$ or $w_{x}<16$; there it climbs
to $1$.

\section{{\normalsize{}Calculation of the bounds (\ref{eq: lam agree upper bound})
and (\ref{eq:  lam agree lower bound})\label{app: calc of bounds}}}

From here on, we will assume $w=w_{x}=w_{p}$ and $k=k_{x}=k_{p}$.

In order to calculate the bounds on $\left\langle \boldsymbol{p}_{\textrm{agree}}\right\rangle $
we will have to find a different way to express $\Pi_{X;0}\Pi_{P;0}\Pi_{X;0}$.
Recalling Eq. (\ref{eq:P_0 inner prod}) and the function (\ref{eq:def of delta})
we now have
\[
\left|\braket{P_{0};m'}{P_{0};m}\right|=\left|\varDelta_{w}\left(\frac{m-m'}{k}\right)\right|=\frac{\sin\left(\pi\frac{m-m'}{k}\right)}{w\sin\left(\pi\frac{m-m'}{d}\right)}.
\]
Observe that the truncated momentum states are orthogonal when the
difference $m-m'$ is an integer number of $k$'s. That is, for any
integers $c$, $c'$ and $n$ the states $\ket{P_{0};ck+n}$ and $\ket{P_{0};c'k+n}$
are orthogonal.

In Eq. (\ref{eq: P0P0P0 in terms of |P_0,m>}) we have derived the
form

\begin{equation}
\Pi_{X;0}\Pi_{P;0}\Pi_{X;0}=\frac{1}{k}\sum_{m=0}^{w-1}\ket{P_{0};m}\bra{P_{0};m}\label{eq:P0P0P0 form recall}
\end{equation}
where $\ket{P_{0};m}\bra{P_{0};m}$ are rank 1 projections. Since
some of these projections are pairwise orthogonal, we can group them
together and express $\Pi_{X;0}\Pi_{P;0}\Pi_{X;0}$ as a smaller sum
of higher rank projections.

In order to do that, let us first assume that $\gamma=w/k$ is a non-zero
integer (we will not need this assumption in general). Then the set
of integers $\left\{ m=0,...,w-1\right\} $ can be partitioned into
$k$ subsets $\Omega_{n}=\left\{ ck+n\,|\,c=0,...,\gamma-1\right\} $
with $n=0,...,k-1$. Thus, we can group up the orthogonal elements
in the sum (\ref{eq:P0P0P0 form recall}) as 
\[
\Pi_{X;0}\Pi_{P;0}\Pi_{X;0}=\frac{1}{k}\sum_{n=0}^{k-1}\sum_{m\in\Omega_{n}}\ket{P_{0};m}\bra{P_{0};m}=\frac{1}{k}\sum_{n=0}^{k-1}\Pi^{\left(n\right)}
\]
where we have introduced the rank $\gamma$ projections 
\[
\Pi^{\left(n\right)}=\sum_{m\in\Omega_{n}}\ket{P_{0};m}\bra{P_{0};m}=\sum_{c=0}^{\gamma-1}\ket{P_{0};ck+n}\bra{P_{0};ck+n}.
\]

When $\gamma=w/k$ is not an integer, the accounting of indices is
more involved. We have to introduce the integer part $g=\left\lfloor \gamma\right\rfloor $
and the remainder part $r=w-k\left\lfloor \gamma\right\rfloor $ of
$\gamma$. As before, we partition the set $\left\{ m=0,...,w-1\right\} $
into subsets 
\[
\Omega_{n}:=\begin{cases}
\left\{ ck+n\,|\,c=0,...,g\right\}  & n<r\\
\left\{ ck+n\,|\,c=0,...,g-1\right\}  & n\geq r
\end{cases}
\]
but now they are not of equal size and the range of $n$ depends on
whether $\gamma\ge1$. When $\gamma\ge1$ then $\left|\Omega_{n}\right|$
is $g+1$ for $n<r$ and $g$ for $n\geq r$. When $\gamma<1$ so
$g=0$ and $r=w$, then $\left|\Omega_{n}\right|=1$ for $n<w$ but
$\left|\Omega_{n}\right|=0$ for $n\geq w$ so we do not need to count
$\Omega_{n}$ for $n\geq w$. Noting that the condition $\gamma\geq1$
is equivalent to $\min\left(k,w\right)=k$ and the condition $\gamma<1$
is equivalent to $\min\left(k,w\right)=w$, we conclude that we only
have to count $\Omega_{n}$ for $n<\min\left(k,w\right)$. Therefore,
for the general $\gamma$ we have
\begin{equation}
\Pi_{X;0}\Pi_{P;0}\Pi_{X;0}=\frac{1}{k}\sum_{n=0}^{\min\left(k,w\right)-1}\sum_{m\in\Omega_{n}}\ket{P_{0};m}\bra{P_{0};m}=\frac{1}{k}\sum_{n=0}^{\min\left(k,w\right)-1}\Pi^{\left(n\right)}\label{eq: P0P0P0 in terms of P^(n)}
\end{equation}
and the projections 
\[
\Pi^{\left(n\right)}=\sum_{m\in\Omega_{n}}\ket{P_{0};m}\bra{P_{0};m}=\sum_{c=0}^{g_{n}-1}\ket{P_{0};ck+n}\bra{P_{0};ck+n}
\]
are now of the rank 
\[
g_{n}=\begin{cases}
g+1 & n<r\\
g & n\geq r.
\end{cases}
\]

Using the new form (\ref{eq: P0P0P0 in terms of P^(n)}), we can re-express
Eq. (\ref{eq: lam_agree in terms of P0P0P0}) as
\begin{equation}
\left\langle \boldsymbol{p}_{\textrm{agree}}\right\rangle =\frac{k^{2}}{d}tr\left[\left(\Pi_{X;0}\Pi_{P;0}\Pi_{X;0}\right)^{2}\right]=\frac{1}{d}\sum_{n,n'=0}^{\min\left(k,w\right)-1}tr\left[\Pi^{\left(n\right)}\Pi^{\left(n'\right)}\right].\label{eq: lam_agree in terms of P^n P^n'}
\end{equation}

\subsection*{The upper bound}

The quantity $tr\left[\Pi^{\left(n\right)}\Pi^{\left(n'\right)}\right]$
is the Hilbert-Schmidt inner product $\left\langle \Pi^{\left(n\right)},\Pi^{\left(n'\right)}\right\rangle $
(also known as Frobenius inner product) of the operators $\Pi^{\left(n\right)}$
and $\Pi^{\left(n'\right)}$. Therefore, it obeys the Cauchy--Schwarz
inequality 
\[
\left|tr\left[\Pi^{\left(n\right)}\Pi^{\left(n'\right)}\right]\right|^{2}=\left|\left\langle \Pi^{\left(n\right)},\Pi^{\left(n'\right)}\right\rangle \right|^{2}\leq\left\langle \Pi^{\left(n\right)},\Pi^{\left(n\right)}\right\rangle \left\langle \Pi^{\left(n'\right)},\Pi^{\left(n'\right)}\right\rangle =tr\left[\Pi^{\left(n\right)}\right]tr\left[\Pi^{\left(n'\right)}\right].
\]
Since the value 
\[
tr\left[\Pi^{\left(n\right)}\Pi^{\left(n'\right)}\right]=\sum_{m\in\Omega_{n}}\sum_{m'\in\Omega_{n'}}\left|\braket{P_{0};m}{P_{0};m'}\right|^{2}
\]
is clearly real and positive, we get 
\[
tr\left[\Pi^{\left(n\right)}\Pi^{\left(n'\right)}\right]\leq\sqrt{tr\left[\Pi^{\left(n\right)}\right]tr\left[\Pi^{\left(n'\right)}\right]}.
\]

The value of $tr\left[\Pi^{\left(n\right)}\right]$ is the rank of
the projection which is either $g$ or $g+1$ so
\[
tr\left[\Pi^{\left(n\right)}\Pi^{\left(n'\right)}\right]\leq g+1.
\]
Therefore, the form of $\left\langle \boldsymbol{p}_{\textrm{agree}}\right\rangle $
in Eq. (\ref{eq: lam_agree in terms of P^n P^n'}) implies that 
\[
\left\langle \boldsymbol{p}_{\textrm{agree}}\right\rangle \leq\frac{1}{d}\sum_{n,n'=0}^{\min\left(k,w\right)-1}\left(g+1\right)=\left(g+1\right)\frac{\min\left(k,w\right)^{2}}{d}.
\]

When $\gamma\geq1$, this upper bound is greater or equal to $1$
because
\[
\left(g+1\right)\frac{\min\left(k,w\right)^{2}}{d}=\left(g+1\right)\frac{k^{2}}{d}\geq\gamma\frac{k^{2}}{d}=w\frac{k}{d}=1
\]
which is not helpful since we already know that $\left\langle \boldsymbol{p}_{\textrm{agree}}\right\rangle \leq1$
for it is a probability. When $\gamma<1$, on the other hand, we have
$g=0$ and so 
\[
\left(g+1\right)\frac{\min\left(k,w\right)^{2}}{d}=\frac{w^{2}}{d}.
\]
Thus, when $\gamma<1$, which translates to $w<k=d/w$ so $w<\sqrt{d}$,
we have the upper bound
\[
\left\langle \boldsymbol{p}_{\textrm{agree}}\right\rangle \leq\frac{w^{2}}{d}.
\]

\subsection*{The lower bound}

We will now focus on the lower bound of the inner product $tr\left[\Pi^{\left(n\right)}\Pi^{\left(n'\right)}\right]$
for the case $\gamma\geq1$ (so $w\geq\sqrt{d}$ and $\min\left(k,w\right)=k$)
and then substitute the result in Eq. (\ref{eq: lam_agree in terms of P^n P^n'}).

Since we are interested in the lower bound, we can simplify the expression
by discarding the terms $c,c'=g$ in the sum
\[
tr\left[\Pi^{\left(n\right)}\Pi^{\left(n'\right)}\right]=\sum_{c=0}^{g_{n}-1}\sum_{c'=0}^{g_{n'}-1}\left|\braket{P_{0};c'k+n'}{P_{0};ck+n}\right|^{2}\geq\sum_{c,c'=0}^{g-1}\left|\braket{P_{0};c'k+n'}{P_{0};ck+n}\right|^{2}.
\]
According to Eq. (\ref{eq:P_0 inner prod}) we have
\[
\left|\braket{P_{0};c'k+n'}{P_{0};ck+n}\right|^{2}=\left|\varDelta_{w}\left(c-c'+\alpha\right)\right|^{2}
\]
where we have introduced the variable $\alpha=\frac{n-n'}{k}$. We
can now identify the sum 
\[
S\left(\alpha\right)=\sum_{c,c'=0}^{g-1}\left|\varDelta_{w}\left(c-c'+\alpha\right)\right|^{2}\,\,\,\,\,\leq tr\left[\Pi^{\left(n\right)}\Pi^{\left(n'\right)}\right]
\]
and focus on lower bounding $S\left(\alpha\right)$ for all possible
$\alpha$.

Since $\left|\varDelta_{w}\left(x\right)\right|^{2}$ is a symmetric
function of $x$ we have
\[
\left|\varDelta_{w}\left(c-c'+\alpha\right)\right|^{2}=\left|\varDelta_{w}\left(-c+c'-\alpha\right)\right|^{2}
\]
and since the values of $c$ and $c'$ are interchangeable in the
sum, we conclude that $S\left(\alpha\right)$ is a symmetric function
of $\alpha$. Therefore, we only need to consider positive $\alpha=\frac{n-n'}{k}$,
and since $n,n'=0,...,k-1$, it takes the values $\alpha=0,\frac{1}{k},...,\frac{k-1}{k}\in\left[0,1\right]$.

Since the summand in $S\left(\alpha\right)$ only depends on the differences
$l=c-c'$, we can simplify the sum
\[
S\left(\alpha\right)=\sum_{l=-g+1}^{g-1}\left(g-\left|l\right|\right)\left|\varDelta_{w}\left(l+\alpha\right)\right|^{2}=\sum_{l=-g+1}^{g-1}\frac{\left(g-\left|l\right|\right)}{w^{2}}\frac{\sin^{2}\left(\pi\left(l+\alpha\right)\right)}{\sin^{2}\left(\pi\left(l+\alpha\right)/w\right)}
\]
where in the last step we substituted the explicit form of $\varDelta_{w}$.
Note that $\sin^{2}\left(\pi\left(l+\alpha\right)\right)=\sin^{2}\left(\pi\alpha\right)$
for integer $l$ and also $\sin^{-2}\left(\frac{\pi\left(l+\alpha\right)}{w}\right)\geq\left(\frac{\pi\left(l+\alpha\right)}{w}\right)^{-2}$
so we get
\begin{align}
S\left(\alpha\right) & \geq\frac{\sin^{2}\left(\pi\alpha\right)}{\pi^{2}}\sum_{l=-g+1}^{g-1}\frac{g-\left|l\right|}{\left(l+\alpha\right)^{2}}.\label{eq: S as sin times s}
\end{align}
We will now focus on evaluating the lower bound of the sum
\begin{equation}
s\left(\alpha\right)=\sum_{l=-g+1}^{g-1}\frac{g-\left|l\right|}{\left(l+\alpha\right)^{2}}.\label{eq: s alpha sum}
\end{equation}

We can rearrange the elements of this sum as follows:

\[
s\left(\alpha\right)=\frac{g}{\alpha^{2}}+\sum_{l=1}^{g-1}\left[\frac{g-l}{\left(l+\alpha\right)^{2}}+\frac{g-l}{\left(l-\alpha\right)^{2}}\right]=\frac{g}{\alpha^{2}}+\sum_{l=1}^{g-1}\left[\frac{l}{\left(g-l+\alpha\right)^{2}}+\frac{l}{\left(g-l-\alpha\right)^{2}}\right]
\]
where in the last step we simply reversed the order of the elements
in the sum. Now we can introduce the auxiliary variables $\beta_{\pm}=g\pm\alpha$,
so
\begin{align}
s\left(\alpha\right) & =\frac{g}{\alpha^{2}}+\sum_{l=1}^{g-1}\left[\frac{l}{\left(l-\beta_{+}\right)^{2}}+\frac{l}{\left(l-\beta_{-}\right)^{2}}\right]=\frac{g}{\alpha^{2}}+\sum_{l=1}^{g-1}\left[\frac{\beta_{+}}{\left(l-\beta_{+}\right)^{2}}+\frac{1}{\left(l-\beta_{+}\right)}+\frac{\beta_{-}}{\left(l-\beta_{-}\right)^{2}}+\frac{1}{\left(l-\beta_{-}\right)}\right]\nonumber \\
 & =\frac{g}{\alpha^{2}}+s_{1}\left(\alpha\right)+s_{2}\left(\alpha\right)\label{eq: s alpha rearanged sum}
\end{align}
where we have identified the sums of harmonic-like series
\[
s_{1}\left(\alpha\right)=\sum_{l=1}^{g-1}\left[\frac{1}{\left(l-\beta_{-}\right)}+\frac{1}{\left(l-\beta_{+}\right)}\right]\,\,\,\,\,\,\,\,\,\,\,\,\,\,\,\,\,\,\,\,\,\,\,\,\,\,s_{2}\left(\alpha\right)=\sum_{l=1}^{g-1}\left[\frac{\beta_{-}}{\left(l-\beta_{-}\right)^{2}}+\frac{\beta_{+}}{\left(l-\beta_{+}\right)^{2}}\right].
\]

Such sums can be evaluated using the polygamma functions \citep{abramowitz1948handbook}
\[
\psi^{\left(j\right)}\left(x\right):=\frac{d^{j}}{dx^{j}}\ln\Gamma\left(x\right)
\]
where $\Gamma$ is the gamma function that interpolates the factorial
for all real (and complex) values. The two key properties of the polygamma
functions that we will need are the recursion and reflection relations
\begin{align}
\psi^{\left(j\right)}\left(1+x\right) & =\psi^{\left(j\right)}\left(x\right)+\left(-1\right)^{j}\frac{j!}{x^{j+1}}\label{eq:polygamma recursion}\\
\psi^{\left(j\right)}\left(1-x\right) & =\left(-1\right)^{j}\psi^{\left(j\right)}\left(x\right)+\left(-1\right)^{j}\pi\frac{d^{j}}{dx^{j}}\cot\left(\pi x\right).\label{eq:polygamma reflection}
\end{align}

For integer $g$ we can expand $\psi^{\left(j\right)}\left(g-x\right)$
for $j=0,1$ using the recursion relation (\ref{eq:polygamma recursion})
to get 
\begin{align*}
\psi^{\left(0\right)}\left(g-x\right) & =\psi^{\left(0\right)}\left(1-x\right)+\sum_{l=1}^{g-1}\frac{1}{l-x}\\
\psi^{\left(1\right)}\left(g-x\right) & =\psi^{\left(1\right)}\left(1-x\right)-\sum_{l=1}^{g-1}\frac{1}{\left(l-x\right)^{2}}.
\end{align*}
Applying the reflection relation (\ref{eq:polygamma reflection})
and rearranging yields
\begin{align}
 & \sum_{l=1}^{g-1}\frac{1}{l-x}=\psi^{\left(0\right)}\left(g-x\right)-\psi^{\left(0\right)}\left(x\right)-\pi\cot\left(\pi x\right)\label{eq: sum as polygamma 1}\\
 & \sum_{l=1}^{g-1}\frac{1}{\left(l-x\right)^{2}}=-\psi^{\left(1\right)}\left(g-x\right)-\psi^{\left(1\right)}\left(x\right)+\frac{\pi^{2}}{\sin^{2}\left(\pi x\right)}.\label{eq: sum as polygamma 2}
\end{align}

Now, using (\ref{eq: sum as polygamma 1}) and recalling that $g-\beta_{\pm}=\mp\alpha$
we can express $s_{1}\left(\alpha\right)$ as 
\[
s_{1}\left(\alpha\right)=\psi^{\left(0\right)}\left(\alpha\right)-\psi^{\left(0\right)}\left(\beta_{-}\right)+\psi^{\left(0\right)}\left(-\alpha\right)-\psi^{\left(0\right)}\left(\beta_{+}\right)
\]
where the trigonometric terms cancel each other out as they are anti-symmetric
and periodic with integer $g$. We can re-express $\psi^{\left(0\right)}\left(\alpha\right)$
and $\psi^{\left(0\right)}\left(-\alpha\right)$ as $\psi^{\left(0\right)}\left(\alpha+1\right)$
using the recursion (\ref{eq:polygamma recursion}) and reflection
relations (\ref{eq:polygamma reflection}) respectively:
\[
\psi^{\left(0\right)}\left(\alpha\right)+\psi^{\left(0\right)}\left(-\alpha\right)=2\psi^{\left(0\right)}\left(\alpha+1\right)+\pi\cot\left(\pi\alpha\right)-\frac{1}{\alpha}.
\]
We can replace $2\psi^{\left(0\right)}\left(\alpha+1\right)$ with
its lower bound $2\psi^{\left(0\right)}\left(1\right)$ on the interval
$0\leq\alpha<1$ as the function $\psi^{\left(0\right)}\left(x\right)$
is monotonically increasing for $0\leq x$. For the same reason we
can also use the bound $\psi^{\left(0\right)}\left(\beta_{\pm}\right)\leq\psi^{\left(0\right)}\left(g+1\right)$
so we end up with the overall lower bound on the sum
\begin{align}
s_{1}\left(\alpha\right) & \geq2\psi^{\left(0\right)}\left(1\right)-2\psi^{\left(0\right)}\left(g+1\right)+\pi\cot\left(\pi\alpha\right)-\frac{1}{\alpha}.\label{eq:s1 LB}
\end{align}

Similarly, using (\ref{eq: sum as polygamma 2}) we can express $s_{2}\left(\alpha\right)$
as
\[
s_{2}\left(\alpha\right)=-\left[\beta_{-}\psi^{\left(1\right)}\left(\alpha\right)+\beta_{+}\psi^{\left(1\right)}\left(-\alpha\right)\right]-\left[\beta_{-}\psi^{\left(1\right)}\left(\beta_{-}\right)+\beta_{+}\psi^{\left(1\right)}\left(\beta_{+}\right)\right]+\frac{\beta_{-}\pi^{2}}{\sin^{2}\left(\pi\beta_{-}\right)}+\frac{\beta_{+}\pi^{2}}{\sin^{2}\left(\pi\beta_{+}\right)}.
\]
Using the recursion (\ref{eq:polygamma recursion}) and reflection
(\ref{eq:polygamma reflection}) relations, we express
\[
-\left[\beta_{-}\psi^{\left(1\right)}\left(\alpha\right)+\beta_{+}\psi^{\left(1\right)}\left(-\alpha\right)\right]=2\alpha\psi^{\left(1\right)}\left(\alpha+1\right)-\frac{\beta_{-}}{\alpha^{2}}-\frac{\beta_{+}\pi^{2}}{\sin^{2}\left(\pi\alpha\right)}\geq-\frac{\beta_{-}}{\alpha^{2}}-\frac{\beta_{+}\pi^{2}}{\sin^{2}\left(\pi\alpha\right)}
\]
where in the last step we have replaced $2\alpha\psi^{\left(1\right)}\left(\alpha+1\right)$
with its lower bound $0$ at $\alpha=0$. Since $\psi^{\left(1\right)}\left(x\right)$
is monotonically decreasing for $0\leq x$ we also use the lower bound
\[
-\left[\beta_{-}\psi^{\left(1\right)}\left(\beta_{-}\right)+\beta_{+}\psi^{\left(1\right)}\left(\beta_{+}\right)\right]\geq-2g\psi^{\left(1\right)}\left(g-1\right).
\]
Thus, the overall lower bound for $s_{2}\left(\alpha\right)$ is
\begin{align}
s_{2}\left(\alpha\right)\geq & -\frac{\beta_{-}}{\alpha^{2}}-\frac{\beta_{+}\pi^{2}}{\sin^{2}\left(\pi\alpha\right)}-2g\psi^{\left(1\right)}\left(g-1\right)+\frac{\beta_{-}\pi^{2}}{\sin^{2}\left(\pi\beta_{-}\right)}+\frac{\beta_{+}\pi^{2}}{\sin^{2}\left(\pi\beta_{+}\right)}\nonumber \\
= & -\frac{\beta_{-}}{\alpha^{2}}-2g\psi^{\left(1\right)}\left(g-1\right)+\frac{\beta_{-}\pi^{2}}{\sin^{2}\left(\pi\alpha\right)}.\label{eq: s2 LB}
\end{align}
where in the last step we have used the fact that $\sin^{2}\left(\pi\beta_{\pm}\right)=\sin^{2}\left(\pi\alpha\right)$.

Combining the lower bounds (\ref{eq:s1 LB}) and (\ref{eq: s2 LB})
into Eqs. (\ref{eq: S as sin times s}), (\ref{eq: s alpha sum}),
(\ref{eq: s alpha rearanged sum}), we get 
\[
S\left(\alpha\right)\geq\,\,g-\frac{2\sin^{2}\left(\pi\alpha\right)}{\pi^{2}}\left[\psi^{\left(0\right)}\left(g+1\right)+g\psi^{\left(1\right)}\left(g-1\right)\right]-\alpha+\frac{2\sin^{2}\left(\pi\alpha\right)}{\pi^{2}}\psi^{\left(0\right)}\left(1\right)+\frac{\sin\left(2\pi\alpha\right)}{2\pi}.
\]

On the interval $0\leq\alpha<1$, the minimum value of 
\[
-\alpha+\frac{2\sin^{2}\left(\pi\alpha\right)}{\pi^{2}}\psi^{\left(0\right)}\left(1\right)+\frac{\sin\left(2\pi\alpha\right)}{2\pi}
\]
 is given by $-\epsilon_{1}\approx-1.005$ and the minimum value of
the coefficient $-\frac{2\sin^{2}\left(\pi\alpha\right)}{\pi^{2}}$
is $-\frac{2}{\pi^{2}}$. With that, we can get rid of the dependence
on $\alpha$: 
\begin{align*}
S\left(\alpha\right) & \geq S_{\min}=g-\frac{2}{\pi^{2}}\left(\psi^{\left(0\right)}\left(g+1\right)+g\psi^{\left(1\right)}\left(g-1\right)\right)-\epsilon_{1}.
\end{align*}

We know that $\psi^{\left(0\right)}\left(x\right)$ is a smooth function
for $x>0$ and it is bounded by \citep{alzer1997some}
\[
\ln x-\frac{1}{x}<\psi^{\left(0\right)}\left(x\right)<\ln x-\frac{1}{2x}
\]
so asymptotically the function $\psi^{\left(0\right)}\left(x+1\right)\sim\ln\left(x+1\right)$
and it converges to $\ln x$ from above. Since $\psi^{\left(1\right)}\left(x\right)=d\psi^{\left(0\right)}\left(x\right)/dx$
then asymptotically $\psi^{\left(1\right)}\left(x\right)\sim\frac{1}{x}$
so the function $x\psi^{\left(1\right)}\left(x-1\right)\sim x/\left(x-1\right)$
and it converges to $1$ from above. Therefore, for any $\epsilon_{2}>0$
there is a $x'>0$ such that for all $x>x'$
\[
\psi^{\left(0\right)}\left(x+1\right)+x\psi^{\left(1\right)}\left(x-1\right)\leq\ln x+1+\epsilon_{2}.
\]
Conveniently choosing $\epsilon_{2}=\frac{\pi^{2}}{2}\left(2-\epsilon_{1}\right)-1$
and solving for $x'$ results in $x'\approx1.722$. Thus, for all
$g\geq2>x'$ we have 
\begin{align*}
S_{\min} & \geq g-\frac{2}{\pi^{2}}\left(\ln g+1+\epsilon_{2}\right)-\epsilon_{1}=g-\frac{2}{\pi^{2}}\ln g-2\\
 & \geq\gamma-\frac{2}{\pi^{2}}\ln\gamma-3
\end{align*}
where the last inequality follows from $g=\left\lfloor \gamma\right\rfloor \geq\gamma-1$
and $\ln g\leq\ln\gamma$.

Recalling that $tr\left[\Pi^{\left(n\right)}\Pi^{\left(n'\right)}\right]\geq S\left(\alpha\right)\geq S_{\min}$
and $\gamma=w/k=w^{2}/d$, we return to the Eq. (\ref{eq: lam_agree in terms of P^n P^n'})
and get the result

\begin{align*}
\left\langle \boldsymbol{p}_{\textrm{agree}}\right\rangle  & =\frac{1}{d}\sum_{n,n'=0}^{k-1}tr\left[\Pi^{\left(n\right)}\Pi^{\left(n'\right)}\right]\geq\frac{k^{2}}{d}S_{\min}\geq\frac{1}{w^{2}/d}\left[w^{2}/d-\frac{2}{\pi^{2}}\ln\left(w^{2}/d\right)-3\right]\\
 & =1-\frac{2}{\pi^{2}}\frac{\ln\left(w^{2}/d\right)+3\pi^{2}/2}{w^{2}/d}.
\end{align*}

\twocolumngrid\bibliography{CGL}

\end{document}

%% file: MathMacros.tex
\global\long\def\ket#1{\left|#1\right\rangle }%

\global\long\def\bra#1{\left\langle #1\right|}%

\global\long\def\braket#1#2{\left.\left\langle #1\right.\,\right|\left.#2\right\rangle }%

\global\long\def\ketbra#1#2{\left|#1\right\rangle \left\langle #2\right|}%

%% file: CGL.bbl
\begin{thebibliography}{27}%
\makeatletter
\providecommand \@ifxundefined [1]{%
 \@ifx{#1\undefined}
}%
\providecommand \@ifnum [1]{%
 \ifnum #1\expandafter \@firstoftwo
 \else \expandafter \@secondoftwo
 \fi
}%
\providecommand \@ifx [1]{%
 \ifx #1\expandafter \@firstoftwo
 \else \expandafter \@secondoftwo
 \fi
}%
\providecommand \natexlab [1]{#1}%
\providecommand \enquote  [1]{``#1''}%
\providecommand \bibnamefont  [1]{#1}%
\providecommand \bibfnamefont [1]{#1}%
\providecommand \citenamefont [1]{#1}%
\providecommand \href@noop [0]{\@secondoftwo}%
\providecommand \href [0]{\begingroup \@sanitize@url \@href}%
\providecommand \@href[1]{\@@startlink{#1}\@@href}%
\providecommand \@@href[1]{\endgroup#1\@@endlink}%
\providecommand \@sanitize@url [0]{\catcode `\\12\catcode `\$12\catcode
  `\&12\catcode `\#12\catcode `\^12\catcode `\_12\catcode `\%12\relax}%
\providecommand \@@startlink[1]{}%
\providecommand \@@endlink[0]{}%
\providecommand \url  [0]{\begingroup\@sanitize@url \@url }%
\providecommand \@url [1]{\endgroup\@href {#1}{\urlprefix }}%
\providecommand \urlprefix  [0]{URL }%
\providecommand \Eprint [0]{\href }%
\providecommand \doibase [0]{http://dx.doi.org/}%
\providecommand \selectlanguage [0]{\@gobble}%
\providecommand \bibinfo  [0]{\@secondoftwo}%
\providecommand \bibfield  [0]{\@secondoftwo}%
\providecommand \translation [1]{[#1]}%
\providecommand \BibitemOpen [0]{}%
\providecommand \bibitemStop [0]{}%
\providecommand \bibitemNoStop [0]{.\EOS\space}%
\providecommand \EOS [0]{\spacefactor3000\relax}%
\providecommand \BibitemShut  [1]{\csname bibitem#1\endcsname}%
\let\auto@bib@innerbib\@empty
\bibitem [{\citenamefont {Busch}\ and\ \citenamefont
  {Shilladay}(2006)}]{busch2006complementarity}%
  \BibitemOpen
  \bibfield  {author} {\bibinfo {author} {\bibfnamefont {P.}~\bibnamefont
  {Busch}}\ and\ \bibinfo {author} {\bibfnamefont {C.}~\bibnamefont
  {Shilladay}},\ }\href {\doibase 10.1016/j.physrep.2006.09.001} {\bibfield
  {journal} {\bibinfo  {journal} {Physics Reports}\ }\textbf {\bibinfo {volume}
  {435}},\ \bibinfo {pages} {1} (\bibinfo {year} {2006})}\BibitemShut {NoStop}%
\bibitem [{\citenamefont {Busch}\ \emph {et~al.}(2007)\citenamefont {Busch},
  \citenamefont {Heinonen},\ and\ \citenamefont {Lahti}}]{busch2007heisenberg}%
  \BibitemOpen
  \bibfield  {author} {\bibinfo {author} {\bibfnamefont {P.}~\bibnamefont
  {Busch}}, \bibinfo {author} {\bibfnamefont {T.}~\bibnamefont {Heinonen}}, \
  and\ \bibinfo {author} {\bibfnamefont {P.}~\bibnamefont {Lahti}},\ }\href
  {\doibase 10.1016/j.physrep.2007.05.006} {\bibfield  {journal} {\bibinfo
  {journal} {Physics reports}\ }\textbf {\bibinfo {volume} {452}},\ \bibinfo
  {pages} {155} (\bibinfo {year} {2007})}\BibitemShut {NoStop}%
\bibitem [{\citenamefont {Heisenberg}(1927)}]{heisenberg1927}%
  \BibitemOpen
  \bibfield  {author} {\bibinfo {author} {\bibfnamefont {W.}~\bibnamefont
  {Heisenberg}},\ }\href@noop {} {\bibfield  {journal} {\bibinfo  {journal} {Z.
  Physik}\ }\textbf {\bibinfo {volume} {43}},\ \bibinfo {pages} {172} (\bibinfo
  {year} {1927})}\BibitemShut {NoStop}%
\bibitem [{\citenamefont {Ozawa}(2003)}]{Ozawa03Universally}%
  \BibitemOpen
  \bibfield  {author} {\bibinfo {author} {\bibfnamefont {M.}~\bibnamefont
  {Ozawa}},\ }\href {\doibase 10.1103/PhysRevA.67.042105} {\bibfield  {journal}
  {\bibinfo  {journal} {Phys. Rev. A}\ }\textbf {\bibinfo {volume} {67}},\
  \bibinfo {pages} {042105} (\bibinfo {year} {2003})}\BibitemShut {NoStop}%
\bibitem [{\citenamefont {Busch}\ \emph {et~al.}(2013)\citenamefont {Busch},
  \citenamefont {Lahti},\ and\ \citenamefont {Werner}}]{Busch13proof}%
  \BibitemOpen
  \bibfield  {author} {\bibinfo {author} {\bibfnamefont {P.}~\bibnamefont
  {Busch}}, \bibinfo {author} {\bibfnamefont {P.}~\bibnamefont {Lahti}}, \ and\
  \bibinfo {author} {\bibfnamefont {R.~F.}\ \bibnamefont {Werner}},\ }\href
  {\doibase 10.1103/PhysRevLett.111.160405} {\bibfield  {journal} {\bibinfo
  {journal} {Phys. Rev. Lett.}\ }\textbf {\bibinfo {volume} {111}},\ \bibinfo
  {pages} {160405} (\bibinfo {year} {2013})}\BibitemShut {NoStop}%
\bibitem [{\citenamefont {Branciard}(2013)}]{branciard2013error}%
  \BibitemOpen
  \bibfield  {author} {\bibinfo {author} {\bibfnamefont {C.}~\bibnamefont
  {Branciard}},\ }\href@noop {} {\bibfield  {journal} {\bibinfo  {journal}
  {Proceedings of the National Academy of Sciences}\ }\textbf {\bibinfo
  {volume} {110}},\ \bibinfo {pages} {6742} (\bibinfo {year}
  {2013})}\BibitemShut {NoStop}%
\bibitem [{\citenamefont {Korzekwa}\ \emph {et~al.}(2014)\citenamefont
  {Korzekwa}, \citenamefont {Jennings},\ and\ \citenamefont
  {Rudolph}}]{korzekwa2014operational}%
  \BibitemOpen
  \bibfield  {author} {\bibinfo {author} {\bibfnamefont {K.}~\bibnamefont
  {Korzekwa}}, \bibinfo {author} {\bibfnamefont {D.}~\bibnamefont {Jennings}},
  \ and\ \bibinfo {author} {\bibfnamefont {T.}~\bibnamefont {Rudolph}},\
  }\href@noop {} {\bibfield  {journal} {\bibinfo  {journal} {Physical Review
  A}\ }\textbf {\bibinfo {volume} {89}},\ \bibinfo {pages} {052108} (\bibinfo
  {year} {2014})}\BibitemShut {NoStop}%
\bibitem [{\citenamefont {Buscemi}\ \emph {et~al.}(2014)\citenamefont
  {Buscemi}, \citenamefont {Hall}, \citenamefont {Ozawa},\ and\ \citenamefont
  {Wilde}}]{buscemi2014noise}%
  \BibitemOpen
  \bibfield  {author} {\bibinfo {author} {\bibfnamefont {F.}~\bibnamefont
  {Buscemi}}, \bibinfo {author} {\bibfnamefont {M.~J.}\ \bibnamefont {Hall}},
  \bibinfo {author} {\bibfnamefont {M.}~\bibnamefont {Ozawa}}, \ and\ \bibinfo
  {author} {\bibfnamefont {M.~M.}\ \bibnamefont {Wilde}},\ }\href@noop {}
  {\bibfield  {journal} {\bibinfo  {journal} {Physical review letters}\
  }\textbf {\bibinfo {volume} {112}},\ \bibinfo {pages} {050401} (\bibinfo
  {year} {2014})}\BibitemShut {NoStop}%
\bibitem [{\citenamefont {Rozema}\ \emph {et~al.}(2015)\citenamefont {Rozema},
  \citenamefont {Mahler}, \citenamefont {Hayat},\ and\ \citenamefont
  {Steinberg}}]{rozema2015note}%
  \BibitemOpen
  \bibfield  {author} {\bibinfo {author} {\bibfnamefont {L.~A.}\ \bibnamefont
  {Rozema}}, \bibinfo {author} {\bibfnamefont {D.~H.}\ \bibnamefont {Mahler}},
  \bibinfo {author} {\bibfnamefont {A.}~\bibnamefont {Hayat}}, \ and\ \bibinfo
  {author} {\bibfnamefont {A.~M.}\ \bibnamefont {Steinberg}},\ }\href {\doibase
  10.1007/s40509-014-0027-1} {\bibfield  {journal} {\bibinfo  {journal}
  {Quantum Studies: Mathematics and Foundations}\ }\textbf {\bibinfo {volume}
  {2}},\ \bibinfo {pages} {17} (\bibinfo {year} {2015})}\BibitemShut {NoStop}%
\bibitem [{\citenamefont {Peres}(2006)}]{peres2006quantum}%
  \BibitemOpen
  \bibfield  {author} {\bibinfo {author} {\bibfnamefont {A.}~\bibnamefont
  {Peres}},\ }\href {\doibase 10.1007/0-306-47120-5} {\emph {\bibinfo {title}
  {Quantum theory: concepts and methods (Chapter 12)}}},\ Vol.~\bibinfo
  {volume} {57}\ (\bibinfo  {publisher} {Springer Science \& Business Media},\
  \bibinfo {year} {2006})\BibitemShut {NoStop}%
\bibitem [{\citenamefont {Kofler}\ and\ \citenamefont
  {Brukner}(2007)}]{kofler2007classical}%
  \BibitemOpen
  \bibfield  {author} {\bibinfo {author} {\bibfnamefont {J.}~\bibnamefont
  {Kofler}}\ and\ \bibinfo {author} {\bibfnamefont {{\v{C}}.}~\bibnamefont
  {Brukner}},\ }\href {\doibase 10.1103/PhysRevLett.99.180403} {\bibfield
  {journal} {\bibinfo  {journal} {Phys. Rev. Lett.}\ }\textbf {\bibinfo
  {volume} {99}},\ \bibinfo {pages} {180403} (\bibinfo {year}
  {2007})}\BibitemShut {NoStop}%
\bibitem [{\citenamefont {Raeisi}\ \emph {et~al.}(2011)\citenamefont {Raeisi},
  \citenamefont {Sekatski},\ and\ \citenamefont {Simon}}]{Raeisi11}%
  \BibitemOpen
  \bibfield  {author} {\bibinfo {author} {\bibfnamefont {S.}~\bibnamefont
  {Raeisi}}, \bibinfo {author} {\bibfnamefont {P.}~\bibnamefont {Sekatski}}, \
  and\ \bibinfo {author} {\bibfnamefont {C.}~\bibnamefont {Simon}},\ }\href
  {\doibase 10.1103/PhysRevLett.107.250401} {\bibfield  {journal} {\bibinfo
  {journal} {Phys. Rev. Lett.}\ }\textbf {\bibinfo {volume} {107}},\ \bibinfo
  {pages} {250401} (\bibinfo {year} {2011})}\BibitemShut {NoStop}%
\bibitem [{\citenamefont {Jeong}\ \emph {et~al.}(2014)\citenamefont {Jeong},
  \citenamefont {Lim},\ and\ \citenamefont {Kim}}]{Jeong14}%
  \BibitemOpen
  \bibfield  {author} {\bibinfo {author} {\bibfnamefont {H.}~\bibnamefont
  {Jeong}}, \bibinfo {author} {\bibfnamefont {Y.}~\bibnamefont {Lim}}, \ and\
  \bibinfo {author} {\bibfnamefont {M.~S.}\ \bibnamefont {Kim}},\ }\href
  {\doibase 10.1103/PhysRevLett.112.010402} {\bibfield  {journal} {\bibinfo
  {journal} {Phys. Rev. Lett.}\ }\textbf {\bibinfo {volume} {112}},\ \bibinfo
  {pages} {010402} (\bibinfo {year} {2014})}\BibitemShut {NoStop}%
\bibitem [{\citenamefont {Rudnicki}\ \emph
  {et~al.}(2012{\natexlab{a}})\citenamefont {Rudnicki}, \citenamefont
  {Walborn},\ and\ \citenamefont {Toscano}}]{Rudnicki_2012a}%
  \BibitemOpen
  \bibfield  {author} {\bibinfo {author} {\bibfnamefont {{\L}.}~\bibnamefont
  {Rudnicki}}, \bibinfo {author} {\bibfnamefont {S.~P.}\ \bibnamefont
  {Walborn}}, \ and\ \bibinfo {author} {\bibfnamefont {F.}~\bibnamefont
  {Toscano}},\ }\href {\doibase 10.1209/0295-5075/97/38003} {\bibfield
  {journal} {\bibinfo  {journal} {{EPL} (Europhysics Letters)}\ }\textbf
  {\bibinfo {volume} {97}},\ \bibinfo {pages} {38003} (\bibinfo {year}
  {2012}{\natexlab{a}})}\BibitemShut {NoStop}%
\bibitem [{\citenamefont {Rudnicki}\ \emph
  {et~al.}(2012{\natexlab{b}})\citenamefont {Rudnicki}, \citenamefont
  {Walborn},\ and\ \citenamefont {Toscano}}]{Rudnicki_2012b}%
  \BibitemOpen
  \bibfield  {author} {\bibinfo {author} {\bibfnamefont {L.}~\bibnamefont
  {Rudnicki}}, \bibinfo {author} {\bibfnamefont {S.~P.}\ \bibnamefont
  {Walborn}}, \ and\ \bibinfo {author} {\bibfnamefont {F.}~\bibnamefont
  {Toscano}},\ }\href {\doibase 10.1103/PhysRevA.85.042115} {\bibfield
  {journal} {\bibinfo  {journal} {Phys. Rev. A}\ }\textbf {\bibinfo {volume}
  {85}},\ \bibinfo {pages} {042115} (\bibinfo {year}
  {2012}{\natexlab{b}})}\BibitemShut {NoStop}%
\bibitem [{\citenamefont {Toscano}\ \emph {et~al.}(2018)\citenamefont
  {Toscano}, \citenamefont {Tasca}, \citenamefont {Rudnicki},\ and\
  \citenamefont {Walborn}}]{toscano2018uncertainty}%
  \BibitemOpen
  \bibfield  {author} {\bibinfo {author} {\bibfnamefont {F.}~\bibnamefont
  {Toscano}}, \bibinfo {author} {\bibfnamefont {D.~S.}\ \bibnamefont {Tasca}},
  \bibinfo {author} {\bibfnamefont {{\L}.}~\bibnamefont {Rudnicki}}, \ and\
  \bibinfo {author} {\bibfnamefont {S.~P.}\ \bibnamefont {Walborn}},\ }\href
  {\doibase 10.3390/e20060454} {\bibfield  {journal} {\bibinfo  {journal}
  {Entropy}\ }\textbf {\bibinfo {volume} {20}},\ \bibinfo {pages} {454}
  (\bibinfo {year} {2018})}\BibitemShut {NoStop}%
\bibitem [{\citenamefont {Ali}\ \emph {et~al.}(2009)\citenamefont {Ali},
  \citenamefont {Das},\ and\ \citenamefont {Vagenas}}]{ali2009discreteness}%
  \BibitemOpen
  \bibfield  {author} {\bibinfo {author} {\bibfnamefont {A.~F.}\ \bibnamefont
  {Ali}}, \bibinfo {author} {\bibfnamefont {S.}~\bibnamefont {Das}}, \ and\
  \bibinfo {author} {\bibfnamefont {E.~C.}\ \bibnamefont {Vagenas}},\ }\href
  {\doibase 10.1016/j.physletb.2009.06.061} {\bibfield  {journal} {\bibinfo
  {journal} {Physics Letters B}\ }\textbf {\bibinfo {volume} {678}},\ \bibinfo
  {pages} {497} (\bibinfo {year} {2009})}\BibitemShut {NoStop}%
\bibitem [{\citenamefont {Hossenfelder}(2013)}]{hossenfelder2013minimal}%
  \BibitemOpen
  \bibfield  {author} {\bibinfo {author} {\bibfnamefont {S.}~\bibnamefont
  {Hossenfelder}},\ }\href {\doibase 10.12942/lrr-2013-2} {\bibfield  {journal}
  {\bibinfo  {journal} {Living Reviews in Relativity}\ }\textbf {\bibinfo
  {volume} {16}},\ \bibinfo {pages} {2} (\bibinfo {year} {2013})}\BibitemShut
  {NoStop}%
\bibitem [{\citenamefont {Ali}\ \emph {et~al.}(2011)\citenamefont {Ali},
  \citenamefont {Das},\ and\ \citenamefont {Vagenas}}]{ali2011proposal}%
  \BibitemOpen
  \bibfield  {author} {\bibinfo {author} {\bibfnamefont {A.~F.}\ \bibnamefont
  {Ali}}, \bibinfo {author} {\bibfnamefont {S.}~\bibnamefont {Das}}, \ and\
  \bibinfo {author} {\bibfnamefont {E.~C.}\ \bibnamefont {Vagenas}},\
  }\href@noop {} {\bibfield  {journal} {\bibinfo  {journal} {Physical Review
  D}\ }\textbf {\bibinfo {volume} {84}},\ \bibinfo {pages} {044013} (\bibinfo
  {year} {2011})}\BibitemShut {NoStop}%
\bibitem [{\citenamefont {Pikovski}\ \emph {et~al.}(2012)\citenamefont
  {Pikovski}, \citenamefont {Vanner}, \citenamefont {Aspelmeyer}, \citenamefont
  {Kim},\ and\ \citenamefont {Brukner}}]{pikovski2012probing}%
  \BibitemOpen
  \bibfield  {author} {\bibinfo {author} {\bibfnamefont {I.}~\bibnamefont
  {Pikovski}}, \bibinfo {author} {\bibfnamefont {M.~R.}\ \bibnamefont
  {Vanner}}, \bibinfo {author} {\bibfnamefont {M.}~\bibnamefont {Aspelmeyer}},
  \bibinfo {author} {\bibfnamefont {M.}~\bibnamefont {Kim}}, \ and\ \bibinfo
  {author} {\bibfnamefont {{\v{C}}.}~\bibnamefont {Brukner}},\ }\href {\doibase
  10.1038/nphys2262} {\bibfield  {journal} {\bibinfo  {journal} {Nature
  Physics}\ }\textbf {\bibinfo {volume} {8}},\ \bibinfo {pages} {393} (\bibinfo
  {year} {2012})}\BibitemShut {NoStop}%
\bibitem [{\citenamefont {Vourdas}(2004)}]{vourdas2004quantum}%
  \BibitemOpen
  \bibfield  {author} {\bibinfo {author} {\bibfnamefont {A.}~\bibnamefont
  {Vourdas}},\ }\href {\doibase 10.1088/0034-4885/67/3/r03} {\bibfield
  {journal} {\bibinfo  {journal} {Reports on Progress in Physics}\ }\textbf
  {\bibinfo {volume} {67}},\ \bibinfo {pages} {267} (\bibinfo {year}
  {2004})}\BibitemShut {NoStop}%
\bibitem [{\citenamefont {Jagannathan}\ \emph {et~al.}(1981)\citenamefont
  {Jagannathan}, \citenamefont {Santhanam},\ and\ \citenamefont
  {Vasudevan}}]{jagannathan1981finite}%
  \BibitemOpen
  \bibfield  {author} {\bibinfo {author} {\bibfnamefont {R.}~\bibnamefont
  {Jagannathan}}, \bibinfo {author} {\bibfnamefont {T.}~\bibnamefont
  {Santhanam}}, \ and\ \bibinfo {author} {\bibfnamefont {R.}~\bibnamefont
  {Vasudevan}},\ }\href {\doibase 10.1007/BF00674253} {\bibfield  {journal}
  {\bibinfo  {journal} {International Journal of Theoretical Physics}\ }\textbf
  {\bibinfo {volume} {20}},\ \bibinfo {pages} {755} (\bibinfo {year}
  {1981})}\BibitemShut {NoStop}%
\bibitem [{\citenamefont {Busch}\ \emph {et~al.}(1996)\citenamefont {Busch},
  \citenamefont {Lahti},\ and\ \citenamefont
  {Mittelstaedt}}]{busch1996quantum}%
  \BibitemOpen
  \bibfield  {author} {\bibinfo {author} {\bibfnamefont {P.}~\bibnamefont
  {Busch}}, \bibinfo {author} {\bibfnamefont {P.~J.}\ \bibnamefont {Lahti}}, \
  and\ \bibinfo {author} {\bibfnamefont {P.}~\bibnamefont {Mittelstaedt}},\
  }\href {\doibase 10.1007/978-3-540-37205-9_3} {\emph {\bibinfo {title} {The
  quantum theory of measurement}}}\ (\bibinfo  {publisher} {Springer},\
  \bibinfo {year} {1996})\BibitemShut {NoStop}%
\bibitem [{\citenamefont {Singh}\ and\ \citenamefont
  {Carroll}(2018)}]{singh2018modeling}%
  \BibitemOpen
  \bibfield  {author} {\bibinfo {author} {\bibfnamefont {A.}~\bibnamefont
  {Singh}}\ and\ \bibinfo {author} {\bibfnamefont {S.~M.}\ \bibnamefont
  {Carroll}},\ }\href@noop {} {\enquote {\bibinfo {title} {Modeling position
  and momentum in finite-dimensional hilbert spaces via generalized clifford
  algebra},}\ } (\bibinfo {year} {2018}),\ \Eprint
  {http://arxiv.org/abs/1806.10134} {arXiv:1806.10134 [quant-ph]} \BibitemShut
  {NoStop}%
\bibitem [{\citenamefont {Poulin}(2005)}]{poulin2005macroscopic}%
  \BibitemOpen
  \bibfield  {author} {\bibinfo {author} {\bibfnamefont {D.}~\bibnamefont
  {Poulin}},\ }\href {\doibase 10.1103/PhysRevA.71.022102} {\bibfield
  {journal} {\bibinfo  {journal} {Physical Review A}\ }\textbf {\bibinfo
  {volume} {71}},\ \bibinfo {pages} {022102} (\bibinfo {year}
  {2005})}\BibitemShut {NoStop}%
\bibitem [{\citenamefont {Abramowitz}\ and\ \citenamefont
  {Stegun}(1972)}]{abramowitz1948handbook}%
  \BibitemOpen
  \bibfield  {author} {\bibinfo {author} {\bibfnamefont {M.}~\bibnamefont
  {Abramowitz}}\ and\ \bibinfo {author} {\bibfnamefont {I.~A.}\ \bibnamefont
  {Stegun}},\ }\href@noop {} {\emph {\bibinfo {title} {Handbook of mathematical
  functions with formulas, graphs, and mathematical tables}}},\ Vol.~\bibinfo
  {volume} {55}\ (\bibinfo  {publisher} {US Government printing office},\
  \bibinfo {year} {1972})\BibitemShut {NoStop}%
\bibitem [{\citenamefont {Alzer}(1997)}]{alzer1997some}%
  \BibitemOpen
  \bibfield  {author} {\bibinfo {author} {\bibfnamefont {H.}~\bibnamefont
  {Alzer}},\ }\href {\doibase /10.1090/S0025-5718-97-00807-7} {\bibfield
  {journal} {\bibinfo  {journal} {Mathematics of computation}\ }\textbf
  {\bibinfo {volume} {66}},\ \bibinfo {pages} {373} (\bibinfo {year}
  {1997})}\BibitemShut {NoStop}%
\end{thebibliography}%
